\documentclass[%
 reprint,
superscriptaddress,
nofootinbib,
 amsmath,amssymb,
 aps,
pra,
floatfix,
]{revtex4-2}
\pdfoutput=1
\usepackage[T1]{fontenc}
\usepackage{amsmath}
\usepackage{graphicx}
\usepackage{adjustbox}
\usepackage{dsfont}
\usepackage{enumitem}
\usepackage{quantikz}
\usepackage{tikz}
\usetikzlibrary{quantikz2, trees, positioning, fit, arrows, decorations.pathreplacing,backgrounds}
\usetikzlibrary{calc}
\providecommand{\tikzsetnextfilename}[1]{}
\usepackage{pgfplots}
\usepackage{amsthm}
\usepackage{tabularx}
\usetikzlibrary{spy}
\usepgfplotslibrary{groupplots}
\pgfplotsset{compat=1.18}
\usepackage{float}
\tikzset{    
    qub/.style = {draw=black, inner sep=3pt,minimum size = 8mm,fill = white, font=\Large},
    tail/.style = {level distance = 8mm},
    last/.style = { circle,inner sep=2pt,minimum size = 8mm, fill = gray!40!white, font=\Large},
    E/.style = {-{Stealth[length=1.2mm]}, fill = white},
    ferms/.style = { circle,inner sep=3pt,minimum size = 6mm, fill = orange!40},
    center/.style = {level distance = 12mm},
    X/.style = {draw=green!70!black, dashdotted, line width=1pt},
    Y/.style = {draw=blue!70!black, densely dashed, line width=1pt},
    Z/.style = {draw=red!80!black, solid, line width=1pt},
    Zc/.style = {draw=black,fill=red!50!white,solid, line width=1pt},
    Yc/.style = {draw=black,fill=blue!40!white,solid, line width=1pt},
    Xc/.style = {draw=black,fill=green!50!white,solid, line width=1pt},
    level distance = 1.6cm,
    edge from parent path={
            (\tikzparentnode.south).. controls +(0,-0.8) and +(0,0.8).. (\tikzchildnode.north)
      },
    s/.style = {inner xsep=-1pt,inner ysep=-1pt}
}
\usepackage{twoopt}
\usepackage[caption=false]{subfig}
\usepackage{booktabs}
\usepackage{array}
\usepackage{adjustbox}
\usepackage{xcolor}
\DeclareRobustCommand{\revised}[2]{\begingroup\color{#2}}
\DeclareRobustCommand{\endrevised}[1]{\endgroup}
\usepackage{silence}
\usepackage{hyperref}

\everymath{\displaystyle}

\newtheorem{theorem}{Theorem}[section]
\newtheorem{lemma}[theorem]{Lemma}

\newcommand{\x}{{\bf x}}

\newcommand{\Ham}{\mathrm{\hat{H}}}

\newcommand{\R}{\mathds{\tilde{R}}}

\newcommand{\cre}[2]{\hat{#1}^{\dagger}_{#2}}
\newcommand{\ann}[2]{\hat{#1}_{#2}}

\newcommand{\mswap}[2]{\hat{\mathrm{M}}_{#1,#2}}

\newcommand{\annq}[2]{#1_{#2}}
\newcommand{\fswapq}[2]{\mathrm{F}_{#1,#2}}
\newcommand{\mswapq}[2]{\mathrm{M}_{#1,#2}}

\newcommand{\TopSquareBracket}[5][]{%
  \draw[thick,#1] let
      \p1=(#2.north), \p2=(#3.north),
      \n1={max(\y1,\y2)}
    in
      (\x1,{\n1+#4}) -- (\x1,{\n1+#4+#5}) --
      (\x2,{\n1+#4+#5}) -- (\x2,{\n1+#4});
}

\DeclareOldFontCommand{\sf}{\normalfont\sffamily}{\mathsf} 
\providecommand{\DIFaddtex}[1]{{\protect\color{black}#1}}
\providecommand{\DIFdeltex}[1]{}
\providecommand{\DIFadd}[1]{\texorpdfstring{\DIFaddtex{#1}}{#1}}
\providecommand{\DIFdel}[1]{}

\providecommand{\DIFaddFL}[1]{\DIFadd{#1}}
\providecommand{\DIFdelFL}[1]{}

\begin{document}


\title{Improving fermionic variational quantum eigensolvers with Majorana swap networks}

\author{D.\,E.\,\surname{Fisher}}
\email{fisher.de19@physics.msu.ru}
\affiliation{%
 Faculty of Physics, M.\,V. Lomonosov Moscow State University, Leninskie Gory 1, Moscow, 119991,
 Russia
}

\author{S.\,A.\,\surname{Fldzhyan}}
\affiliation{%
 Russian Quantum Center, Bolshoy bul'var 30 building 1, Moscow, 121205, Russia
}
\affiliation{%
 Faculty of Physics, M.\,V. Lomonosov Moscow State University, Leninskie Gory 1, Moscow, 119991,
 Russia
}

\author{D.\,V.\,\surname{Minaev}}
\affiliation{%
 Faculty of Physics, M.\,V. Lomonosov Moscow State University, Leninskie Gory 1, Moscow, 119991,
 Russia
}

\author{S.\,S.\,\surname{Straupe}}
\affiliation{Sber Quantum {\color{black}Technologies} Center, Kutuzovski prospect 32, Moscow, 121170, Russia}
\affiliation{%
 Russian Quantum Center, Bolshoy bul'var 30 building 1, Moscow, 121205, Russia
}
\affiliation{%
 Faculty of Physics, M.\,V. Lomonosov Moscow State University, Leninskie Gory 1, Moscow, 119991,
 Russia
}

\author{M.\,Yu.\,\surname{Saygin}}
\affiliation{Sber Quantum {\color{black}Technologies} Center, Kutuzovski prospect 32, Moscow, 121170, Russia}
\affiliation{%
 Faculty of Physics, M.\,V. Lomonosov Moscow State University, Leninskie Gory 1, Moscow, 119991,
 Russia
}

\begin{abstract}

Simulating computationally hard fermionic systems is a promising application of quantum
computing. However, mapping nonlocal fermionic operators to qubits often produces deep circuits,
rendering such simulations impractical on near-term hardware. We introduce two Majorana swap
network compilation strategies for variational quantum eigensolvers that reduce circuit depth
and two-qubit gate count. First, we develop a cyclic compilation algorithm that localizes all
two-particle interaction terms in a general fermionic Hamiltonian
{\revised{1}{black}containing up to $\mathcal{O}(M^4)$ such terms using only
$\mathcal{O}(M^3)$ auxiliary Majorana-swap transpositions, where $M$\endrevised{1}}
is the number of fermionic modes. {
  \revised{1}{black}Here, the cubic scaling refers to auxiliary
routing; a complete UCCGSD ansatz still contains $\mathcal{O}(M^4)$ double-excitation
rotations.\endrevised{1}} Second, we design a Majorana swap network for the
 $k$-UpCCGSD variational ansatz, which is already more compact than UCCGSD.
 In this setting,
our network yields constant-factor reductions
 of approximately $50$\%
 in circuit depth
 and
$20$\%
 in two-qubit gate count
 under all-to-all connectivity. For the more restricted
$2\times N$ connectivity, the reductions are larger --- about $55$\%
 in circuit depth
 and
$40$\%
 in gate count. These structural improvements are accompanied by improved robustness in
numerical noise
  \DIFadd{simulations }{\revised{2}{black}on the small molecular instances tested.\endrevised{2} }

\end{abstract}


\maketitle


\section{Introduction}
\label{sec:Intro}

Accurate simulation of correlated fermionic systems underpins progress in quantum chemistry,
materials design, and condensed-matter physics~\cite{corfer1,ModelSystem,SimElectonicStructure}.
While classical electronic-structure methods have achieved remarkable successes, their cost
typically scales unfavorably with system size, reflecting the exponential growth of the
underlying Hilbert space~\cite{CI_method}. Quantum computers promise polynomial-resource
simulations in principle~\cite{QPE,QuantumAlgorithms,FTQCchemistry}, but realizing this promise
on noisy, connectivity-limited devices remains challenging.

A key obstacle is the encoding of fermionic operators into \DIFadd{the} qubit Hilbert space. Standard
mappings such as Jordan–Wigner (JW)~\cite{JW} and Bravyi–Kitaev (BK)~\cite{BK} produce nonlocal
Pauli strings whose support grows with the number of fermionic modes, inflating circuit depth
and    \DIFadd{the number of entangling} gates, and exacerbating error accumulation on Noisy
Intermediate-Scale Quantum (NISQ) hardware~\cite{NISQ_alg}. Numerous strategies aim to mitigate
this nonlocality: alternative encodings (e.g., tree-based and task-based
mappings)~\cite{Wang_2021,Wang_2023,TT, bonsai}, ancilla-assisted compact
encodings~\cite{SBK,compact_mapping,maj_loop,lexic}, and structured gate scheduling via swap
networks (SNs)~\cite{SWAP}. Yet, for widely used variational approaches such as the variational
quantum eigensolver (VQE)~\cite{VQE} with Unitary Coupled Cluster (UCC)
ansatzes~\cite{gener_UCC,Universal_UCC}, the cumulative overhead from nonlocal fermionic
operators can still be significant.

Swap networks have been used to enforce effective locality under JW by bringing interacting
modes next to each other    \DIFadd{using} fermionic swap gates (FSWAP)~\cite{SWAP,LinearDepth}.
This work adopts the Majorana approach: each fermionic mode is represented by two involutive
Majorana operators obeying simple anticommutation relations, providing a natural way to
represent them as elements of the Pauli group. We introduce \emph{Majorana swap} gates (MSWAP)
--- operations that implement signed permutations of two Majoranas --- and show how to assemble
them into \emph{Majorana swap networks} (MSNs). Compared with FSWAPs, MSWAPs offer finer-grained
control over operator support and also admit compact decompositions when mapped to qubits.

We focus on two widely used variants of UCC ansatz. The first is the generalized
singles-and-doubles UCCGSD, which provides high expressivity at the expense of many parameters
and gates. The second is $k$-UpCCGSD, which restricts double excitations to spatially paired
orbitals and is commonly repeated in $k$ layers to recover expressivity with modest overhead
 while preserving the expectation value of the total spin squared
operator~\cite{gener_UCC}. We will show that the $k$-UpCCGSD structure aligns naturally with a
$2\times N$ qubit layout consistent with existing topologies \cite{Sycamore, Willow,
Experiment2xN},
  \DIFadd{in which 
 }$\alpha$ and $\beta$ orbitals
  \DIFadd{are
placed} in separate rows.

Our main contributions are:
\begin{enumerate}
    \item \textbf{Optimized single- and double-excitation decompositions.} We provide gate
 constructions for double-excitation rotations
 that reduce depth and
        two-qubit
  \DIFadd{gate count} compared with existing
        implementations~\cite{EffChemCirc}, and that integrate cleanly with
 the
        $2\times N$ geometry studied here.
    \item \textbf{Cyclic permutation for UCCGSD with
  {\revised{3}{black}\(\mathcal{O}(M^3)\)\endrevised{3}}
        auxiliary-routing
 overhead on all-to-all connectivity.} We present a permutation
        algorithm that preserves fermionic symmetry and guarantees local implementation of all
 double-excitation
  {\revised{3}{black}supports using only \(\mathcal{O}(M^3)\) MSWAP transpositions,
        where \(M\) is the number of fermionic modes. This improves upon the
        \(\mathcal{O}(M^4)\) auxiliary-swap overhead\endrevised{3}} of prior
        approaches~\cite{SWAP}.
  \revised{3}{black}This changes the routing overhead, not the
        number of variational rotations: for the complete UCCGSD ansatz, the circuit still
        contains \(\mathcal{O}(M^4)\) double-excitation rotations after the local terms are
        exposed. Although this reduction does not change the asymptotic rotation count,
        it can still reduce constants and becomes relevant when swap and measurement
        overheads are decisive~\cite{TILLY20221,alfonso2025chemical,qubit_adapt_vqe,overlap_adapt_vqe,ramoa2025reducing,co_adapt_swap_penalty}.\endrevised{3}
     \item \textbf{MSN circuits for \texorpdfstring{$k$-UpCCGSD}{k-UpCCGSD}.} Inspired by the
        fermionic swap network (FSN)~\cite{SWAP,LinearDepth}, we design MSN-based compilations
        that localize the required excitation rotations while maximizing parallel execution.
 In our construction, the observed savings arise not only from the reduced
        swap overhead of the MSN, but also from fusing optimized single- and double-excitation
        rotations directly into the MSWAP layers. For the benchmarked $1$-UpCCGSD circuits, this
        combined synthesis reduces circuit depth by about $50$\% and the two-qubit gate count by
        about $20$\% relative to the FSN baseline under all-to-all connectivity. On $2\times N$
        layouts, to which the $k$-UpCCGSD is naturally matched, the reductions increase to about
        $55$\% in depth and $40$\% in two-qubit gate count.
 \item \textbf{Noise-aware evaluation.} Using representative depolarizing and correlated
        error channels, as well as
 superconducting- and ion-device noise
        models, we quantify the sensitivity of different compilation strategies for
 the $k$-UpCCGSD ansatz. \revised{4}{black}MSNs generally show lower or near-lowest energy susceptibility among the
        fixed-baseline alternatives considered here, with channel-dependent exceptions discussed
        below.\endrevised{4}
 \end{enumerate}

Section~\ref{sec:preliminaries} reviews background on fermionic encodings, Majorana operators,
and variational ansatzes. Section~\ref{sec:MSWAP} introduces MSWAP transformations and contrasts
them with FSWAP. Section~\ref{sec:implementations_excitations} presents gate-level
decompositions for single and double excitations. Section~\ref{sec:cyclic_permut} develops the
cyclic-permutation algorithm for UCCGSD. Section~\ref{sec:majorana_SWAP} applies MSNs to the
\(k\)-UpCCGSD ansatz with \DIFadd{a \(2\times N\) connectivity constraint.} Section~\ref{sec:numerical_study} reports numerical depth and gate-count benchmarks
and noise simulations. Section~\ref{sec:conclusion} concludes with implications for noise-aware
ansatz design and future extensions.


\section{Preliminary Details}\label{sec:preliminaries}

In the formalism of second quantization, the quantum state of    \DIFadd{$M$} fermionic modes is described
by creation $\cre{a}{p}$ and annihilation $\ann{a}{p}$ operators, where the index $p$ (   \DIFadd{$1\leq
p\leq M$} ) labels the single-particle fermionic orthonormal states. These operators satisfy the
canonical anticommutation relations:
    \begin{equation}
        \{\cre{a}{p},\cre{a}{q}\}=0, \quad
        \{\ann{a}{p}, \ann{a}{q}\}=0, \quad 
        \{\cre{a}{p}, \ann{a}{q}\}=\delta_{pq}.
    \end{equation} 

These anticommutation relations prevent a fully local encoding, since any two operators must
act nontrivially on at least one common qubit. However, for systems with conserved particle
number, only operator products containing equal numbers of creation and annihilation operators
are allowed. \DIFadd{Therefore} only even-length products of creation and annihilation operators
are physically allowed; we refer to such products as \emph{physical operators}. The physical
operators form the foundation for representing fermionic states and constructing fermionic
Hamiltonians used in simulation.

 \subsection{Variational quantum eigensolver}

We    \DIFadd{consider VQE} as a promising approach for leveraging
NISQ-era quantum devices~\cite{VQA,VQA_Error_correct,VQE}. The variational algorithms aim to
approximate the ground state of a target Hamiltonian by preparing a parameterized quantum state
and minimizing the expected energy. As in classical variational methods, VQEs restrict the
optimization to a polynomially parameterized subspace of the exponentially large Hilbert space.
This subspace is constructed by applying a parameterized unitary circuit
$U(\boldsymbol{\theta})$ to a fixed reference state $\ket{\psi_0}$:
    \begin{equation}
        \ket{\psi(\boldsymbol{\theta})} = U(\boldsymbol{\theta}) \ket{\psi_0}.
    \end{equation}
The corresponding cost function is defined as
    \begin{equation}
        L(\boldsymbol{\theta}) = \bra{\psi(\boldsymbol{\theta})} \Ham
        \ket{\psi(\boldsymbol{\theta})},
    \end{equation}
where $\Ham$ is the Hamiltonian of interest.

VQEs offer two main advantages. First, they can efficiently prepare highly entangled quantum
states using native unitary dynamics. Second, they introduce an inductive bias: by tailoring the
ansatz structure to salient features of the target state, they can make the optimization
landscape more favorable.

Beyond quantum simulation, VQEs have been explored for diverse tasks, including integer
factorization~\cite{factoring}, solving nonlinear equations~\cite{nonlinear_equation}, and
quantum machine learning~\cite{QML,QML_chal}. A wide range of VQE ansatz architectures has been
proposed~\cite{HEA, HVA, gener_UCC, adaptvqe1, adaptvqe2}, trading off expressiveness,
trainability, and hardware compatibility. In this work we focus on the \emph{unitary
coupled-cluster} (\emph{UCC}) family~\cite{gener_UCC}, which extends classical coupled
cluster~\cite{CI_method} by exponentiating anti-Hermitian excitation operators. A general UCC
ansatz takes the form:
    \begin{equation}\label{eq:ansatzUCC1}
        \ket{\psi(\mathbf{s}, \mathbf{d})} = \exp\bigg[ \sum_{(p,q)\in S}\!\!\! s_{pq} \hat{T}_{pq}  + \!\!\!\!\!\!\sum_{(p,q,r,s)\in D}\!\!\! \!\!\!d_{pqrs} \hat{T}_{pqrs} \bigg] \ket{\psi_0},
    \end{equation}
where $\mathbf{s}$ and $\mathbf{d}$ are the sets of real-valued parameters $s_{pq}$ and
$d_{pqrs}$, and $S$ and $D$ are sets of index tuples corresponding to
 particle-number- and spin-projection-preserving
 single and double excitations, respectively. In
Eq.~\eqref{eq:ansatzUCC1}, the excitation operators are given by:
    \begin{align}
        \hat{T}_{pq} &= \cre{a}{p} \ann{a}{q} - \cre{a}{q} \ann{a}{p},\nonumber \\
        \hat{T}_{pqrs} &= \cre{a}{p} \cre{a}{q} \ann{a}{r} \ann{a}{s} - \cre{a}{s} \cre{a}{r} \ann{a}{q} \ann{a}{p}.
    \end{align}

To make the ansatz implementable on quantum hardware, Eq.~\eqref{eq:ansatzUCC1} is typically
Trotterized \cite{Trotter1959}. The resulting circuit, up to operator reordering, takes the
form:
    \begin{equation}
        \ket{\psi(\mathbf{s}, \mathbf{d})} =\!\!\! \prod_{(p,q)\in S} \!\!\!\hat{S}_{pq}(s_{pq}) \!\!\!\prod_{(p,q,r,s)\in D}\!\!\! \hat{D}_{pqrs}(d_{pqrs}) \ket{\psi_0}.
    \end{equation}
\DIFadd{Here} $\hat{S}_{pq}(s_{pq})$ and $\hat{D}_{pqrs}(d_{pqrs})$ denote the single and double
excitation rotations, respectively:
    \begin{equation}\label{eq:trot_ucc}
        \hat{S}_{pq}(s) = e^{s\hat{T}_{pq}},\quad \hat{D}_{pqrs}(d) =e^{d\hat{T}_{pqrs}}.
    \end{equation}

For the UCCGSD ansatz, the excitation index sets are given by:
\begin{align}
    S_G &=
    \{ (p_{\sigma},q_{\sigma}) \mid
    1\le p<q\le N,\;
    \sigma\in\{\alpha,\beta\} \}, \\
    D_G &=
    \{ (p_{\sigma},q_{\tau},r_{\sigma},s_{\tau}) \mid
    p_{\sigma},q_{\tau},r_{\sigma},s_{\tau}\text{ are distinct} \}.
\end{align}
where    \DIFadd{$1 \leq p, q, r, s \leq N$} label spatial orbitals and $\alpha, \beta$ denote  \DIFadd{the} two spin
components.  \DIFadd{Thus the corresponding number of spin orbitals and fermionic modes is
$M=2N$.
}

For this ansatz, we introduce a \emph{cyclic-permutation algorithm} that localizes all
{\revised{5}{black}$\mathcal{O}(M^4)$ generalized double-excitation supports using only $\mathcal{O}(M^3)$
auxiliary MSWAP transpositions, thereby reducing the auxiliary routing overhead.\endrevised{5}}

Although UCCGSD can be highly expressive and performs well in small-scale studies, its resource
demands grow rapidly with system size.
 Deep randomly initialized UCC-type
circuits can exhibit barren-plateau behavior.

To mitigate these issues, we also examine a more resource-efficient variant,
 $k$-UpCCGSD, which restricts double excitations to spatially paired orbitals and repeats them
with independent parameters across
 $k$ layers. In the spin-complemented form
considered here, the same single-excitation amplitude is applied to the $\alpha$ and $\beta$
channels, which is compatible with the total-spin symmetry of the paired
construction~\cite{gener_UCC}. The double-excitation sector
 therefore includes only excitations
of the form:
    \begin{equation}\label{eq:Dup}
        D_{\text{Up}} = \left\{ (p_\alpha, p_\beta, q_\alpha, q_\beta) \mid 1 \leq p < q \leq N
        \right\}.
    \end{equation}
  \DIFadd{The ansatz} can be expressed as
  \begin{multline}\label{eq:real_ansatzUCC}
         \ket{\psi(\mathbf{s}, \mathbf{d})} = \prod_{l=1}^k  \Bigg[
        \Bigg(\prod_{\textcolor{black}{p<q}} \hat{S}_{p_\alpha q_\alpha}(s^{(l)}_{pq}) \hat{S}_{p_\beta q_\beta}(s^{(l)}_{pq})\Bigg) \\
         \times \Bigg(\prod_{\textcolor{black}{p<q}} \hat{D}_{p_\alpha p_\beta
         q_\alpha q_\beta}(d^{(l)}_{pq})\Bigg)
        \Bigg] \ket{\psi_0}.
    \end{multline} 

It offers a favorable trade-off between expressiveness and circuit complexity, enabling
chemically accurate results with significantly fewer parameters
  \DIFadd{~} \cite{gener_UCC}.

\subsection{Majorana operators and the Pauli group}\label{subsec:maj}

To implement the unitary evolution described by~\eqref{eq:trot_ucc} on a quantum computer, the
second quantization operators must first be mapped onto qubit operators. This step is
complicated by the anticommutation relations of fermionic operators, which often lead to
nonlocal qubit representations and deep quantum circuits.

Several fermion-to-qubit mappings have been developed to address this
challenge~\cite{compact_mapping,BK,SBK,loc_ham,loc_spin,maj_loop}. These mappings aim to reduce
operator nonlocality, often at the cost of introducing auxiliary qubits. In fully connected
systems, these approaches may require up to twice as many qubits~\cite{loc_ham,loc_spin,SBK} to
achieve representations where physical operators (those composed of equal numbers of creation
and annihilation operators) act nontrivially on only $\mathcal{O}(1)$ qubits. We refer to these
operators    \DIFadd{acting nontrivially on $w$ qubits as}\emph{   \DIFadd{$w$}-local}.

Two principal paradigms exist for constructing fermionic mappings~\cite{tapering}:
\begin{itemize}
    \item \textbf{First quantization}, which encodes fermionic Fock states directly. This approach can reduce the number of required qubits using symmetries, but it often results in deeper and more intricate circuit constructions;
    \item \textbf{Second quantization}, which encodes creation and annihilation operators through Pauli operators. This method is more widely used in near-term quantum computing due to its compatibility with standard hardware and \DIFadd{lower} circuit depth.
\end{itemize}

In this work, we adopt the second quantization framework and focus on fermion-to-qubit mappings
based on \emph{Majorana operators}, represented using \emph{ternary tree (TT)} structures, which
we explain below~\cite{TT, Vlasov}. Majorana operators $\hat{\gamma}_j$ are defined as
    \begin{equation}
        \begin{aligned}
            &\hat{\gamma}_{2p-1}       = \hat{a}_p + \hat{a}_p^\dagger, 
            &\qquad 
            &\hat{\gamma}_{2p}         = i(\hat{a}_p^\dagger - \hat{a}_p), \\
            &\hat{a}_p                 = \frac{\hat{\gamma}_{2p-1} + i\hat{\gamma}_{2p}}{2}, 
            &\qquad 
            &\hat{a}_p^\dagger         = \frac{\hat{\gamma}_{2p-1} - i\hat{\gamma}_{2p}}{2}, \\
        \end{aligned}
    \end{equation}
that obey the anticommutation relations:
    \begin{equation}\label{eqn:anticommut}
        \{\hat{\gamma}_p, \hat{\gamma}_q\} = 2\delta_{pq}. 
    \end{equation}
The operators $\hat{\gamma}_j$ are Hermitian and unitary, and their algebra resembles that of
Hermitian elements of the Pauli group \DIFadd{$G_M = \{\iota X, \iota Y, \iota Z, \iota I\}^{\otimes
M}$}, where $\iota$ is a fourth root of unity. The corresponding Pauli strings can be used to
construct qubit-space representations of Majorana operators. In this representation, the number
of fermionic modes    \DIFadd{$M$ equals} the number of qubits in the system.

A valid mapping is obtained by associating each Majorana operator $\hat{\gamma}_j$ with a Pauli
string $\gamma_j$ such that the resulting set $\{\gamma_j\}$ is mutually anti-commuting.
Henceforth, we omit the hat symbol when referring to the Pauli representation. It has been shown
that pairwise anticommutation is a necessary and sufficient condition for a valid
fermion-to-qubit mapping~\cite{fermion_map_theory}. It turns out that stricter conditions, such
as algebraic independence, are not required~\cite{bonsai,fermihedral}.

The TT formalism provides a compact and scalable way to construct such mutually anti-commuting
sets, while preserving qubit locality, which is an essential requirement for low-depth circuit
implementations.

In general, constructing mappings where all physical operators are strictly
$\mathcal{O}(1)$-local is not possible without ancillary qubits. However, this limitation can be
circumvented by using \emph{SNs}, which dynamically reorder fermionic
modes~\cite{SWAP,SWAP_QAOA}. SNs can be implemented efficiently in certain mappings, such as the
JW mapping~\cite{SWAP,LinearDepth}, and allow locality to be maintained throughout the circuit.

In this work, we use SNs to support constant-locality implementations of physical operators
without the need for ancilla qubits. These networks act as switching layers that transform
between different representations of fermionic or Majorana operators, enabling ansatz
implementation under realistic connectivity constraints.

\subsection{Ternary tree representation}
\label{subsec:tt}

By employing Pauli strings from the group    \DIFadd{$G_M$ to represent $M$} fermionic modes, one can
construct an exponential number of valid fermion-to-qubit mappings --- on the order of
   \DIFadd{$N_{\text{all}} = \mathcal{O}(2^{M^2})$.} A significant subset of these mappings can be captured
using \emph{TTs} --- graphical structures in which each path from the root to a leaf encodes a
Majorana operator~\cite{TT,bonsai,Vlasov}. Several well-known mappings, such as JW, BK and the
parity mapping~\cite{parity}, admit TT representations~\cite{Vlasov}.

\begin{figure}[hbtp!]
    \centering
    \def\twidth{0.45}
  \subfloat[Mapping via full TT. $\gamma_1 = X_1X_2$, $\gamma_5 = Y_1Z_3$]{
        \resizebox{0.46\linewidth}{!}{\includegraphics{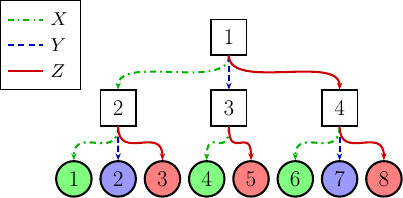}}
        \label{fig:tt_ex}
} \hfill
  \subfloat[JW mapping via TT. $\gamma_1 = X_1$, $\gamma_5 = Z_1Z_2X_3$]{
        \resizebox{0.36\linewidth}{!}{\includegraphics{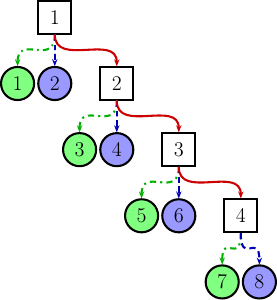}}
        \label{fig:jw_ex}
  }
     \caption{\label{fig:tt1}\DIFaddFL{Ternary-tree (TT) representations of fermion-to-qubit mappings. 
Panel (a) shows a generic TT, and panel (b) shows the Jordan--Wigner (JW) TT. 
White square nodes denote qubits; circular leaves denote Majorana operators. 
A Majorana Pauli string is obtained by following the path from the root to the corresponding leaf.}}
    \end{figure}

In  \DIFadd{the} TT representation shown in Fig.~\ref{fig:tt1}, square nodes represent qubits, while circular
leaf nodes correspond to Majorana operators. To obtain the Pauli string associated with a given
Majorana operator, one traverses the tree from the root to the corresponding leaf. The Pauli
operator ($X$, $Y$, or $Z$) assigned to each qubit along the path is determined by the type of
edge taken. The anticommutation relations between Majorana operators  \DIFadd{in} Eq.~\eqref{eqn:anticommut}
are guaranteed by construction, since any two distinct paths differ nontrivially at exactly one
qubit --- the node where the paths diverge    \DIFadd{--- ensuring} the correct algebraic structure.

As illustrated in Fig.~\ref{fig:tt1}, different mappings yield varying degrees of operator
locality. Among them, the JW mapping exhibits the highest variability. In this encoding,
Majorana operators are mapped to Pauli strings as follows:
    \begin{equation}
        \begin{aligned}
            \gamma_{2p - 1} &= Z_1 Z_2 \cdots Z_{p-1}\, X_{p}, \\
            \gamma_{2p}     &= Z_1 Z_2 \cdots Z_{p-1}\, Y_{p},
        \end{aligned}
    \end{equation}
which generally results in highly nonlocal operators. However, products involving adjacent
modes can remain local. For example, $\gamma_5 \gamma_6 = i Z_3$ acts only on a single qubit.
Throughout the text, operators written without the hat symbol $\hat{\cdot}$ refer to their
qubit-space representations, as determined by the current mapping that includes all previously
applied permutations.

This degenerate locality is a key feature that enables the efficient use of swap networks in the
JW mapping, allowing for dynamic reordering of fermionic modes and substantially reducing
  circuit depth and gate count.



\section{SWAP Transformations for Fermionic and Majorana Modes}\label{sec:MSWAP}

SNs rely fundamentally on gates that reorder modes while preserving fermionic anticommutation
relations. In this section, we review two classes of such gates --- FSWAP and MSWAP    \DIFadd{--- focusing
} in particular on the structural benefits of MSWAP gates for operator localization and circuit
simplification.

\subsection{Fermionic swap}

\begin{figure}[htbp!]
    \centering
    \includegraphics[width=0.6\linewidth]{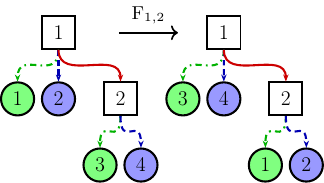}
    \hfill
    \includegraphics[width=1.0\linewidth]{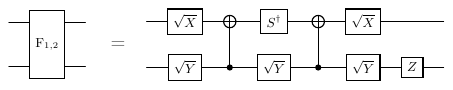}%
     \caption{\DIFaddFL{Two-$\mathrm{CX}$ implementation of the adjacent fermionic swap $\fswapq{1}{2}$ in the JW encoding. 
The gate exchanges two neighboring fermionic modes while preserving the fermionic signs.}}
    \label{fig:FSWAP}
\end{figure}

The FSWAP gate exchanges two fermionic modes while maintaining fermionic antisymmetry. It is
defined as
    \begin{equation}
    \hat{F}_{p,q} = 1 + \cre{a}{p} \ann{a}{q} + \cre{a}{q} \ann{a}{p} - \cre{a}{p} \ann{a}{p} - \cre{a}{q} \ann{a}{q},
    \end{equation}
where indices $p$ and $q$ designate the modes to be exchanged. This gate satisfies the identities:
    \begin{equation}
        \hat{F}_{p,q} \, \ann{a}{p} \, \hat{F}_{p,q}^\dagger = \ann{a}{q}, \quad
        \hat{F}_{p,q} \, \ann{a}{q} \, \hat{F}_{p,q}^\dagger = \ann{a}{p}.
    \end{equation}

When applied to adjacent modes in the JW encoding, the FSWAP gate admits a particularly
efficient implementation using only two CX gates, which is illustrated in Fig.~\ref{fig:FSWAP}.
 For reference, the ordinary qubit SWAP gate decomposes as
\begin{equation}
    \mathrm{SWAP}_{p,q}=\mathrm{CX}_{p,q}\,\mathrm{CX}_{q,p}\,\mathrm{CX}_{p,q}.
\end{equation}
Thus   in the JW encoding, the fermionic
phase structure allows it to be implemented more compactly.

\subsection{Majorana swap}

\begin{figure}[htbp!]
    \centering
    \includegraphics[width=0.6\linewidth]{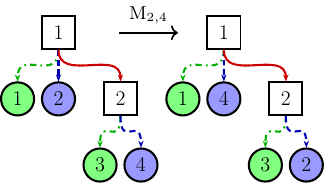}
    \hfill
    \includegraphics[width=1.0\linewidth]{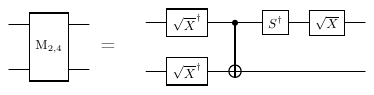}%
    \caption{\DIFaddFL{Implementation of the MSWAP gate $\mswapq{2}{4}=e^{\pi\gamma_2\gamma_4/4}=e^{i\pi X_1Y_2/4}$ in the JW encoding.}}
    \label{fig:MSWAP}
\end{figure}

As discussed in Sec.~\ref{subsec:maj}, fermionic operators $\hat{a}_p$ and $\hat{a}_p^\dagger$
can be equivalently expressed in terms of Majorana operators $\hat{\gamma}_{2p-1}$ and
$\hat{\gamma}_{2p}$. The MSWAP gate acts directly on these operators and performs a signed
permutation. For a pair of Majorana modes $\hat{\gamma}_i$ and $\hat{\gamma}_j$, the MSWAP
operator is defined as
    \begin{equation}
        \hat{M}_{i,j} = e^{\frac{\pi}{4} \hat{\gamma}_i \hat{\gamma}_j},
    \end{equation}
   \DIFadd{which} acts as
    \begin{equation}
        \hat{M}_{i,j} \hat{\gamma}_i \hat{M}_{i,j}^\dagger = -\hat{\gamma}_j, \quad
        \hat{M}_{i,j} \hat{\gamma}_j \hat{M}_{i,j}^\dagger = \hat{\gamma}_i.
    \end{equation}

When the involved Majorana operators correspond to adjacent modes in the JW encoding, the MSWAP
gate can be implemented using a single two-qubit Clifford operation, as illustrated in
Fig.~\ref{fig:MSWAP}.

  \begingroup
\revised{6}{black}
While both MSWAP and FSWAP gates implement related fermionic reorderings, MSWAP gates act on
individual Majorana operators and therefore provide finer control over the intermediate operator
structure. In particular, a fermionic swap can be assembled from two MSWAP gates. For example,
\begin{equation}
\begin{aligned}
&\mswap{1}{3}\mswap{2}{4}\cre{a}{1}
\mswap{2}{4}^{\dagger}\mswap{1}{3}^{\dagger}
= -\cre{a}{2},\\
&\mswap{1}{3}\mswap{2}{4}\cre{a}{2}
\mswap{2}{4}^{\dagger}\mswap{1}{3}^{\dagger}
= \cre{a}{1}.
\end{aligned}
\end{equation}
Thus, \(\mswap{1}{3}\mswap{2}{4}\) implements the fermionic exchange up to
mode-dependent signs. The sign is not global phase: it can either be absorbed
into the signs of subsequent rotation angles or corrected by the intra-mode operation
\(\mswap{1}{2}^2\), which corresponds
to a single-qubit \(Z\) gate in the JW encoding.

In the constructions below, however, we use a different transformation,
\(\mswap{3}{2}\mswap{1}{4}\), because its intermediate Majorana ordering exposes a
more useful modified JW encoding for localizing excitation operators:
\begin{equation}
\begin{aligned}
&\mswap{3}{2}\mswap{1}{4} \cre{a}{1}
\mswap{1}{4}^{\dagger}\mswap{3}{2}^{\dagger}
= -i\cre{a}{2},\\
&\mswap{3}{2}\mswap{1}{4} \cre{a}{2}
\mswap{1}{4}^{\dagger}\mswap{3}{2}^{\dagger}
= -i\cre{a}{1}.
\end{aligned}
\label{eq:FSWAP_from_MSWAP}
\end{equation}
The remaining single-mode phases can be removed by an additional MSWAP layer, with
appropriate orientations of \(\mswap{1}{2}\) and \(\mswap{3}{4}\). Equivalently, they can
be kept as part of the tracked signed Majorana ordering and accounted for in later
MSWAP transpositions and excitation-rotation implementations. Here, the signed ordering specifies the current permutation of the Majorana operators together with their associated $\pm 1$ signs. This is the point used
below: MSWAPs are not used merely as a resource-equivalent replacement of FSWAPs,
but as a way to pass through intermediate Majorana encodings with shorter Pauli
representations of the targeted excitations.
\endrevised{6}
\endgroup


\section{Excitation Implementations}\label{sec:implementations_excitations}

This section introduces gate decompositions for local single- and double-excitation rotations
defined in Eq.~\eqref{eq:trot_ucc}, which are necessary for constructing variational circuits.
We propose a method for implementing these excitation rotations integrated with fermionic swap
operations via MSWAP gates, which substantially reduces both the gate count and the circuit
depth.

\subsection{Standard single- and double-excitation rotations}

\begin{figure}[hbtp!]
\centering
  \includegraphics{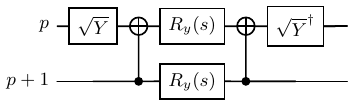}
 \caption{\label{fig:single_exc}\DIFaddFL{Example decomposition of the local single-excitation rotation $S_{p,p+1}$ between adjacent fermionic modes in the JW mapping.}}
\end{figure}

\begin{figure}[tbp]
     \centering
  \resizebox{\linewidth}{!}{\includegraphics{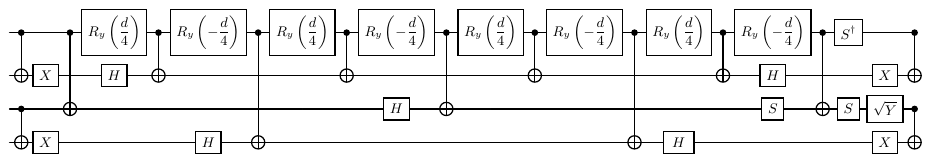}}
     \caption{\DIFaddFL{Yordanov decomposition of a local four-mode double-excitation rotation in the standard JW encoding, Eq.~\eqref{eq:double_jw}.}}
    \label{fig:yordanov}
  \end{figure}

Excitation operators are generally nonlocal unless all fermionic modes involved are adjacent in
the JW mapping. In what follows, we consider this adjacency condition.

 {\revised{7}{black}For compactness, we denote
$\Gamma_{i_1\ldots i_m}\equiv \gamma_{i_1}\cdots\gamma_{i_m}$.\endrevised{7}}
In the JW encoding, a single-excitation rotation corresponds to the sum of two two-qubit Pauli
rotations:
\begin{multline}
S_{p,p+1}(s) = \exp\left[ \frac{s}{2}\left(\Gamma_{2p-1,2p+1} + \Gamma_{2p,2p + 2}\right) \right] \\
= \exp\left[ \frac{i s}{2}\left(X_p Y_{p+1} - Y_p X_{p+1}\right) \right].
\end{multline}
This operator can be implemented with two $\mathrm{CX}$ gates, as shown in
Fig.~\ref{fig:single_exc}.

Double-excitation rotations correspond to eight distinct Pauli strings:
\begingroup
  \revised{7}{black}
The standard JW double-excitation rotation can then be written as
\begin{align}
&D_{1234}(d)
=
\exp\bigg[
\frac{id}{8}
\Big(
    -\Gamma_{2467} +\Gamma_{1367} -\Gamma_{1457} +\Gamma_{1468}
\nonumber\\
&\qquad
    -\Gamma_{2357} +\Gamma_{2368} +\Gamma_{1358} -\Gamma_{2458}
\Big)
\bigg]=
\nonumber\\
&=
\exp\bigg[
\frac{id}{8}
\Big(
    \mathrm{XYXX} +\mathrm{YXXX} +\mathrm{YYYX} +\mathrm{YYXY}
\nonumber\\
&\qquad
    -\mathrm{XXYX}-\mathrm{XXXY}-\mathrm{YXYY} -\mathrm{XYYY}
\Big)
\bigg].
\label{eq:double_jw}
\end{align}
\endrevised{7}
 \endgroup

In this context, mode indices are omitted for clarity. This expression corresponds to the JW
representation of a fermionic double-excitation acting on four consecutive modes.

   {\revised{8}{black}A 13-$\mathrm{CX}$ implementation of this operator was proposed by Yordanov et
al.~\cite{EffChemCirc} and is now a standard CNOT-efficient reference circuit for double
qubit-excitation rotations~\cite{qubit_adapt_vqe,magoulas2023cnot}. We use this circuit as a
known literature baseline\endrevised{8}}  (see Fig.~\ref{fig:yordanov}).

\subsection{   \DIFadd{Single-} and    \DIFadd{double-excitation rotations} in  \DIFadd{the} modified encoding}
\label{sec:modif_exc}

\begin{figure}[hbtp!]
    \centering
  \includegraphics{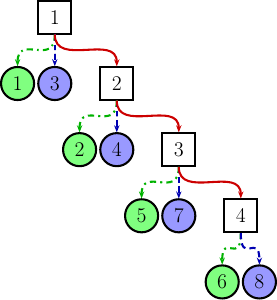}
     \caption{\DIFaddFL{Example of the modified JW encoding obtained after the input MSWAP layer $U_{\mathrm{in}}$ in Eq.~\eqref{eq:uin}. }}
    \label{fig:modified_jw}
\end{figure}

\begin{figure}[hbtp!]
    \centering
      \includegraphics{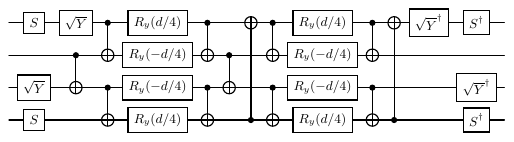}
     \caption{\DIFaddFL{Compact implementation of the modified-encoding double-excitation rotation in Eq.~\eqref{eq:opt_double}. 
The four-qubit core uses 12 $\mathrm{CX}$ gates.}}
    \label{fig:double_opt}
\end{figure}

\begingroup
\revised{9}{black}
Here, we use MSWAP gates to access modified JW encodings in which excitation
operators can have shorter Pauli support. We first apply an input MSWAP layer,
synthesize the excitation rotation in the resulting modified JW encoding, and
then either complete the corresponding fermionic reordering with an output
MSWAP layer or continue in the current signed ordering of Majorana operators.
In the latter case, the Majorana labels and signs are propagated to the
subsequent transformations. This is the mechanism by which excitation rotations
are fused with the swap-network structure.

Suppose first that we apply two MSWAP gates to the four-qubit JW mapping:
\begin{equation}
\label{eq:uin}
    U_{\mathrm{in}}=\mswapq{3}{2}\mswapq{7}{6}.
\end{equation}
The resulting mapping is shown in Fig.~\ref{fig:modified_jw}. With the
convention
\begin{equation}
\mswapq{i}{j}\gamma_i\mswapq{i}{j}^{\dagger}=-\gamma_j,
\qquad
\mswapq{i}{j}\gamma_j\mswapq{i}{j}^{\dagger}=\gamma_i,
\end{equation}
the action of \(U_{\mathrm{in}}\) on the relevant Majorana operators is
\begin{equation}
\label{eq:uin_majorana_table}
\begin{array}{rcl@{\qquad}rcl}
\gamma_1 &\mapsto& \gamma_1,
&
\gamma_5 &\mapsto& \gamma_5,
\\
\gamma_2 &\mapsto& \gamma_3,
&
\gamma_6 &\mapsto& \gamma_7,
\\
\gamma_3 &\mapsto& -\gamma_2,
&
\gamma_7 &\mapsto& -\gamma_6,
\\
\gamma_4 &\mapsto& \gamma_4,
&
\gamma_8 &\mapsto& \gamma_8.
\end{array}
\end{equation}

Substituting Eq.~\eqref{eq:uin_majorana_table} into the standard JW expressions,
the single- and double-excitation rotations in the modified encoding become
\begin{equation}
S_{12}(s)
=
\exp\left[
    \frac{s}{2}\left(-\Gamma_{12}+\Gamma_{34}\right)
\right]
=
\exp\left[
    \frac{is}{2}\left(-Z_1+Z_2\right)
\right],
\label{eq:opt_single}
\end{equation}
\begin{align}
D_{1234}(d)
&=
\exp\bigg[
\frac{id}{8}
\Big(
    -\Gamma_{3467}-\Gamma_{1267}
    +\Gamma_{1456}+\Gamma_{1478}
\nonumber\\
&\qquad
    +\Gamma_{2356}+\Gamma_{2378}
    -\Gamma_{1258}-\Gamma_{3458}
\Big)
\bigg]
\nonumber\\
&=
\exp\bigg[
\frac{id}{8}
\Big(
    \mathrm{IZXX}+\mathrm{ZIXX}
    +\mathrm{YYZI}+\mathrm{YYIZ}
\nonumber\\
&\qquad
    -\mathrm{XXZI}-\mathrm{XXIZ}
    -\mathrm{ZIYY}-\mathrm{IZYY}
\Big)
\bigg].
\label{eq:opt_double}
\end{align}

Equation~\eqref{eq:opt_single} shows that the single-excitation rotation becomes
a product of two single-qubit phase rotations, with the relative signs absorbed
into the rotation angles. Equation~\eqref{eq:opt_double} contains eight distinct
Pauli strings, ordered according to the corresponding terms in the standard JW
double-excitation expression. 

For these double-excitation rotations, we use the compact decomposition shown in
Fig.~\ref{fig:double_opt}. It consists of 12 \(\mathrm{CX}\) gates and achieves
a \(40\%\) reduction in depth compared with the Yordanov decomposition. The
resulting four-qubit core embeds efficiently in the \(2\times 2\) cells of the
\(2\times N\) geometry considered later.
\endrevised{9}
 \endgroup
 
After implementing the excitation-rotation layer, we append a second layer of MSWAP gates,
\begin{equation}
U_{out}=\mswapq{1}{4}\mswapq{5}{8}.
\end{equation}
   {\revised{10}{black}Together with Eq.~\eqref{eq:FSWAP_from_MSWAP}, these two MSWAP layers realize an FSWAP-like
reordering up to the trackable phases discussed above. This enables more efficient
implementations of excitation rotations by integrating them with the swap-network structure.\endrevised{10}}

Furthermore, we propose two MSN constructions (Secs.~\ref{sec:cyclic_permut} and
\ref{sec:majorana_SWAP}) that require fewer swap operations than FSN. Combined with the
techniques above, these constructions further reduce the two-qubit gate count and circuit depth.


\section{Cyclic Permutation Algorithm for Hamiltonian Compilation}\label{sec:cyclic_permut}

\begin{figure*}[hbtp!]
        \centering
          \resizebox{\linewidth}{!}{\includegraphics{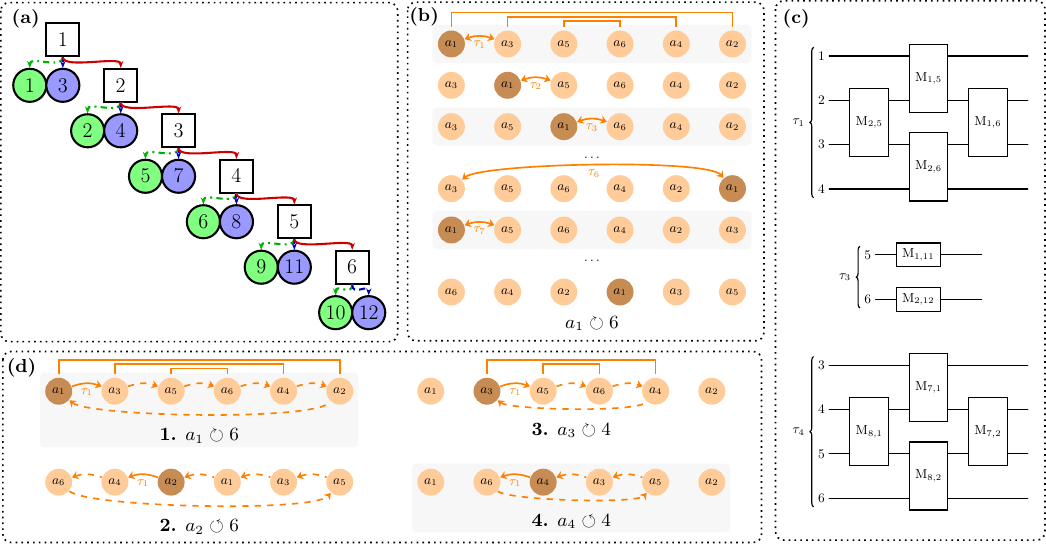}}
         \caption{\label{fig:cycl}{\revised{11}{black}Cyclic permutation algorithm for localizing four-mode supports.
(a) Fermion-to-qubit mapping represented as a TT.
(b) Fermionic modes grouped into local pairs; the selected mode \(a_1\) is swept through the active list by transpositions \(\tau_i\).
(c) Circuit-level realization of these transpositions by local MSWAP-based ordering steps.
(d) Alternating cyclic sweeps over progressively smaller active lists. The procedure exposes all UCCGSD double-excitation supports as unions of two local fermionic pairs using \(\mathcal{O}(M^3)\) auxiliary MSWAP transpositions, while the full UCCGSD pool still contains \(\mathcal{O}(M^4)\) excitation rotations.
\endrevised{11}}}
    \end{figure*}

\begingroup
\revised{12}{black}
This section presents a cyclic MSWAP-routing procedure for localizing four-mode qubit supports, including the supports of double-excitation operators in the UCCGSD ansatz. The procedure uses \(\mathcal{O}(M^3)\) auxiliary MSWAP transpositions to expose the \(\mathcal{O}(M^4)\) possible four-mode supports in a compact modified-JW form. This scaling refers only to the auxiliary Majorana-routing stage: the complete UCCGSD ansatz still contains \(\mathcal{O}(M^4)\) variational double-excitation rotations. Compactness refers to the support in the current fermionic/Majorana ordering and does not by itself imply adjacency on a restricted hardware layout.
\endrevised{12}
\endgroup

\subsection{Cyclic permutation algorithm}

{\revised{13}{black}
We propose a cyclic permutation algorithm to efficiently transform each two-particle interaction term --- that is, a product of two creation and two annihilation operators --- into a compact four-qubit operator. The idea of the algorithm is illustrated in Fig.~\ref{fig:cycl}. We start from the modified JW encoding with reordered Majorana operators, shown in Fig.~\ref{fig:cycl}(a):
\begin{equation}    
\label{eq:2xnmap}
    \begin{aligned}
           \annq{\gamma}{4p - 3}  &= Z_1 \ldots Z_{2p - 2} X_{2p-1},\\
           \annq{\gamma}{4p - 1}  &= Z_1 \ldots Z_{2p - 2} Y_{2p-1},\\
           \annq{\gamma}{4p - 2}  &= Z_1 \ldots Z_{2p - 1} X_{2p},\\
           \annq{\gamma}{4p}      &= Z_1 \ldots Z_{2p - 1} Y_{2p}.
    \end{aligned}
\end{equation}
This encoding is obtained by applying the MSWAP pattern of Eq.~\eqref{eq:uin} pairwise. Each fermionic mode is then distributed over two qubits. As shown in Fig.~\ref{fig:cycl}(b), the ordering naturally groups the modes into local fermionic pairs whose support spans two adjacent qubits. At each stage, we will arrange the fermionic modes which will be permuted into the so-called active list:
\begin{equation}
L=(\ell_0,\ell_1,\ldots,\ell_{2K-1}).
\end{equation}
For the active list, local fermionic pairs correspond to the reflection about the midpoint,
\begin{equation}
P_i(L)=\{\ell_i,\ell_{2K-1-i}\},\qquad i=0,\ldots,K-1.
\end{equation}
Therefore, a double-excitation support becomes compact whenever its four fermionic modes can be written as the union of two distinct local pairs, \(P_i(L)\cup P_j(L)\).
\endrevised{13}}

\begingroup
\revised{13}{black}
The procedure consists of cyclic sweeps in which one selected mode is moved through the active list. This sweep is illustrated in Fig.~\ref{fig:cycl}(b), while the corresponding MSWAP-level transpositions are shown in Fig.~\ref{fig:cycl}(c). We denote such a sweep by
\begin{equation}
    a_i\circlearrowleft 2K
    \qquad \text{or} \qquad
    a_i\circlearrowright 2K,
\end{equation}
where the arrow indicates the sweep direction and \(2K\) is the current active-list size. For \(2K\) active fermionic modes, one sweep of a selected mode requires
\begin{equation}
    T_{\mathrm{sweep}}(K)=K(2K-1)
\end{equation}
local MSWAP transpositions. During the sweep, the local pair containing the selected mode is combined with every other local pair. Hence every four-mode operator containing the selected mode obtains compact qubit support at least once. Appendix~\ref{sec:cycl_appendix} gives the constructive proof.
\endrevised{13}
\endgroup

{\revised{13}{black}
Full localization is achieved by applying this sweep procedure sequentially to different modes. Fig.~\ref{fig:cycl}(d) shows the alternating sweep pattern
\begin{equation}
    a_1\circlearrowleft 6,\qquad
    a_2\circlearrowright 6,\qquad
    a_3\circlearrowleft 4,\qquad
    a_4\circlearrowright 4,\ldots.
\end{equation}
Here \(a_1\circlearrowleft 6\) means that mode \(a_1\) is swept anticlockwise through an active list of six fermionic modes, while \(a_2\circlearrowright 6\) denotes the opposite sweep direction, which restores the paired encoding before the active subset is reduced. The subsequent sweeps \(a_3\circlearrowleft 4\), \(a_4\circlearrowright 4\), and so on follow the same alternating pattern on smaller active lists. This sequential application localizes all four-mode supports without ancillary qubits by systematically processing the active modes while compensating for ordering changes.
\endrevised{13}}

{\revised{14}{black}
Starting from \(M=2N\) fermionic modes, the total number of auxiliary transpositions required by the algorithm is
\begin{equation}
    \begin{aligned}
        T_{\mathrm{cyc}}(N)
        &=
        \sum_{K=3}^{N} 2K(2K-1)  \\
        &=
        \frac{4}{3}N^3 + N^2 - \frac{1}{3}N - 14  \\
        &=
        \frac{1}{6}M^3 + \frac{1}{4}M^2 - \frac{1}{6}M - 14
        =
        \mathcal{O}(M^3).
    \end{aligned}
    \label{eq:cyclic_count}
\end{equation}
The sum starts at \(K=3\), because when only four active modes remain, the single remaining four-mode support is already the union of the two local pairs.
\endrevised{14}}

\subsection{Implementation on a linear qubit layout}

\begin{figure*}[hbtp!]
    \centering
    \subfloat[MSN]{
    \label{fig:msn_up}
    \resizebox{0.48\linewidth}{!}{
    \includegraphics{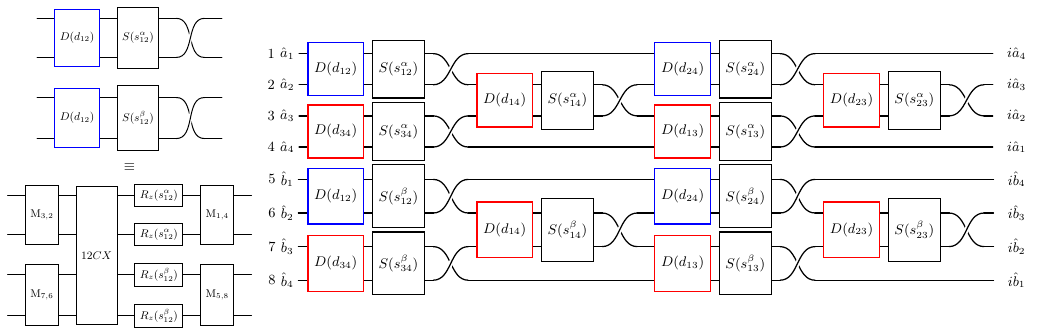}
    }
   }\
    \subfloat[FSN]{
    \label{fig:fsn_up}
    \resizebox{0.48\linewidth}{!}{\includegraphics{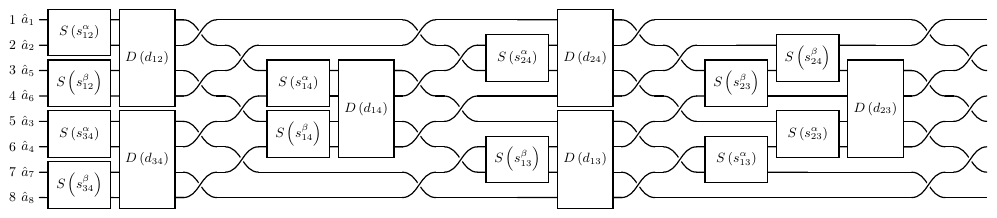}}
   }
    \caption{\label{fig:msn}
    \textbf{Circuit synthesis for an 8-qubit 1-UpCCGSD ansatz}.
    (a) Synthesis using the MSN. The upper and lower halves represent
    $\alpha$- and $\beta$-spin orbitals, respectively. Same-colored
    double-excitation blocks form a single four-qubit rotation. The swap
    operations are fused with the excitation rotations via MSWAPs, as shown
    in the inset on the left. The factors $i$ at the output indicate the
    tracked phases.
    (b) Synthesis using the FSN, where each crossing corresponds to an
    FSWAP.}
    \end{figure*}

\begingroup
\revised{15}{black}
We also indicate how the compact operators exposed by the cyclic MSWAP-routing procedure can be applied on a linear qubit layout. At any step with \(2K\) active fermionic modes, the current ordering defines \(K\) local fermionic pairs. Let \(P_a\) be the pair containing the selected mode \(a\), and let \(P_j\) be another active pair. The double-excitation supports involving \(a\) have the compact form
\begin{equation}
    P_a\cup P_j,\qquad P_j\neq P_a.
\end{equation}
With all-to-all connectivity, the corresponding four-qubit rotation can be applied directly. On a linear layout, compact support in the Majorana ordering does not necessarily imply that the four qubits are geometrically adjacent. One can instead keep \(P_a\) as a two-qubit block and move it along the line until it reaches each \(P_j\). Exchanging two neighboring two-qubit blocks requires at most four ordinary qubit SWAPs. Therefore, one pass of \(P_a\) through the active list uses \(\mathcal{O}(K)\) ordinary qubit SWAPs while exposing \(\mathcal{O}(K)\) four-qubit excitation rotations. The extra hardware-routing cost is thus \(\mathcal{O}(1)\) per applied double-excitation rotation. For the complete \(\mathcal{O}(M^4)\) UCCGSD pool on a linear layout, these ordinary qubit SWAPs give an additional \(\mathcal{O}(M^4)\) hardware-routing overhead on top of the \(\mathcal{O}(M^3)\) MSWAP-routing cost.
\endrevised{15}
\endgroup

For comparison, an explicit swap-network construction for localizing all double-excitation terms was previously proposed in Ref.~\cite{SWAP}. That construction works naturally on linear connectivity and offers better opportunities for parallel execution but can lead to significant overhead in auxiliary SWAP gates due to its recursive building. In contrast, the present cyclic construction is designed to reduce the number of auxiliary MSWAP transpositions.

\begingroup
\revised{16}{black}
We note that the \(\mathcal{O}(M^3)\) auxiliary MSWAP scaling is not only achieved by the cyclic permutation, but is optimal up to constant factors for this JW-localization task. Indeed,
for four fermionic modes ordered as \(i_1<i_2<i_3<i_4\), the product of a four-Majorana term
contains parity $Z$ only on the intervals \((i_1,i_2)\) and \((i_3,i_4)\). These parities
disappear precisely when \(i_1,i_2\) form one adjacent fermionic pair and \(i_3,i_4\) form
another. Thus compact four-mode supports in this ordering are exactly unions of two adjacent
fermionic pairs.

A local MSWAP transposition can change only the adjacent-pair structure involving the transposed
modes. With \(M\) modes, each changed pair can be combined with only
\(\mathcal{O}(M)\) other pairs, so one transposition can expose at most \(\mathcal{O}(M)\) new
compact four-mode supports. Since the UCCGSD pool contains \(\mathcal{O}(M^4)\) such supports, any
local-transposition procedure needs \(\Omega(M^3)\) auxiliary SWAPs in the worst case.
\endrevised{16}
\endgroup

Although the cyclic algorithm reduces the auxiliary-swap overhead for UCCGSD, the full ansatz is still dominated by its \(\mathcal{O}(M^4)\) double-excitation rotations. The resulting circuits are therefore too large for the density-matrix noisy simulations considered below. We next specialize the MSWAP construction to the more compact \(k\)-UpCCGSD ansatz.



\section{Majorana swap network for \texorpdfstring{$k$-UpCCGSD}{k-UpCCGSD}}
\label{sec:majorana_SWAP}
\DIFadd{
As outlined above, the $k$-UpCCGSD ansatz employs unitary evolution
generated by the restricted set of excitation operators defined in
Eq.~\eqref{eq:real_ansatzUCC}. In the spin-complemented form used here,
the singles are applied with the same amplitude in the $\alpha$ and
$\beta$ sectors, while the doubles excite paired $\alpha\beta$ electrons
between spatial orbitals. These generators preserve both $\hat S_z$ and
the total-spin symmetry generated by $\hat S^2$. This spin-sector
structure naturally motivates a $2\times N$ architecture, in which the
upper and lower rows encode the $\alpha$- and $\beta$-spin orbitals,
respectively.
}

\begin{figure}[hbtp!]
    \centering
    \includegraphics{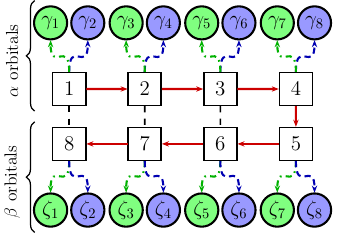}
     \caption{\label{fig:device_connectivity}\DIFaddFL{Example fermion-to-qubit mapping on a $2\times4$ lattice. The two rows encode $\alpha$- and $\beta$-spin orbitals; Majorana operators $\gamma$ and $\zeta$ label the two spin sectors. Solid edges show the JW tree/order used to define the fermionic mapping, while dashed edges show the nearest-neighbor physical couplings allowed in the $2\times N$ device model.}}
\end{figure}

A key structural feature of the UpCCGSD ansatz is that each double-excitation operator acts on a
pair of spin orbitals corresponding to the same spatial orbital. To preserve these pairings
throughout the circuit, we apply synchronized permutations across both rows of the $2\times N$
lattice. These permutations are realized using a \textit{2-complete swap
network}~\cite{SWAP,LinearDepth}—a structured sequence of swap layers that ensures every pair of
fermionic modes becomes adjacent at least once during the circuit execution.

This property is illustrated in Figs.~\ref{fig:msn_up} and \ref{fig:fsn_up}, where each wire
represents a fermionic mode and each crossing indicates a point where the corresponding modes
are brought into adjacency. This enables the efficient implementation of all required two-mode
fermionic operators as local operators after    \DIFadd{an appropriate} layer.

Fig.~\ref{fig:msn_up} depicts an MSN tailored for the 8-qubit $1$-UpCCGSD ansatz. The circuit
alternates between swap and excitation layers. In our construction, MSWAP gates are fused with
double- and single-excitation rotations, as discussed in Sec.~\ref{sec:modif_exc}. As noted, this
integration reduces circuit overhead by embedding excitation operations directly within the swap
layers. Double excitations are realized as localized four-qubit gates, with
same-colored blocks in the two spin sectors representing parts of the
same gate.

 {\revised{17}{black}
After completing the MSN shown in Fig.~\ref{fig:msn_up}, the fermionic
operators in each spin sector appear in the reversed logical order, up
to phase factors generated by the oriented MSWAP gates. These fermionic
phases follow from the tracked signed Majorana ordering and can either
be incorporated into the final measurement basis or removed using
single-qubit $Z$ and $S$ gates. The factors $i$ shown at the output of
Fig.~\ref{fig:msn_up} therefore record the resulting phase convention
and do not represent additional gates.
\endrevised{17}}

By contrast, the FSN proposed in Ref.~\cite{SWAP} and shown in Fig.~\ref{fig:fsn_up} also
   \DIFadd{realizes} a 2-complete network; however, it performs extra permutations between orbitals of
opposite spin, which results in twice as many FSWAP operations. In FSN, single excitations
require two CX gates, the same as FSWAPs. As a result, FSN exhibits reduced parallelism and
increased circuit depth compared to MSNs.


\section{Numerical study }\label{sec:numerical_study}

\begin{figure*}[htbp!]
    \centering
      \resizebox{\linewidth}{!}{\includegraphics{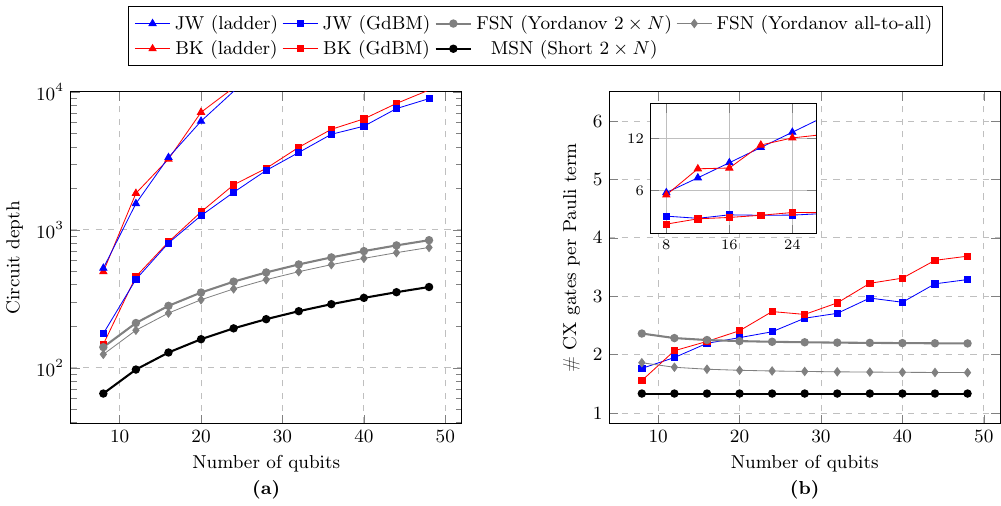}}
     \caption{\revised{18}{black}Resource scaling for compiled $1$-UpCCGSD circuits. 
Panel (a) shows the transpiled circuit depth in $\{\mathrm{CX},U_3\}$ gates, and panel (b) shows the total number of $\mathrm{CX}$ gates divided by the number of Pauli-string rotations in the ansatz which is independent of the circuit compilation method. 
The proposed MSN-based method is shown with black lines. 
These curves are structural compilation benchmarks as a function of qubit number, not noisy VQE simulations.\endrevised{18}}
    \label{fig:compare_kup}
    \end{figure*}

\begin{figure*}[htbp!]
    \centering
      \resizebox{\linewidth}{!}{\includegraphics{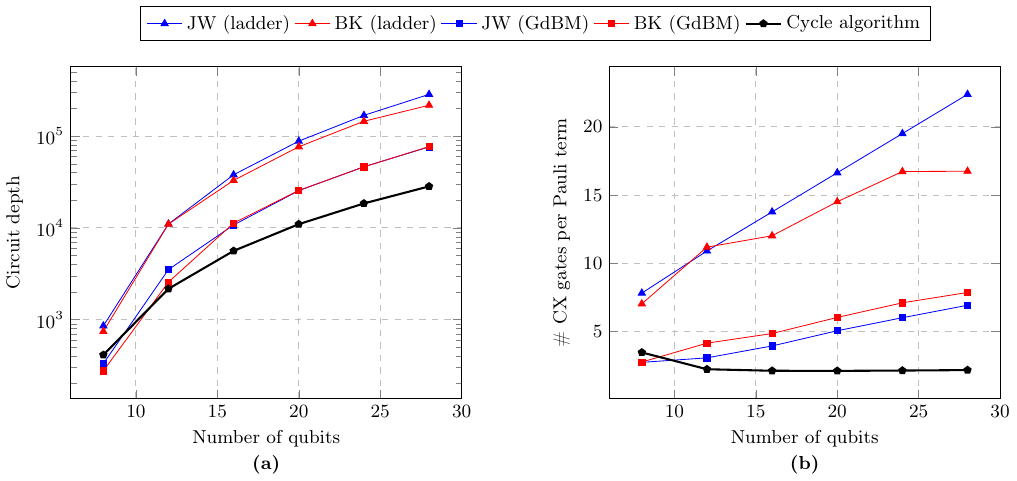}}
     \caption{\revised{19}{black}Resource scaling for compiled UCCGSD circuits. 
Panel (a) shows the transpiled circuit depth in $\{\mathrm{CX},U_3\}$ gates, and panel (b) shows the total number of $\mathrm{CX}$ gates divided by the number of Pauli-string rotations in the ansatz. The black line marks the proposed cyclic-permutation compilation routing. \endrevised{19}}
    \label{fig:compare_ucc}
    \end{figure*}

\subsection{Evaluation of circuit depth and gate count}

To demonstrate the capabilities of the proposed constructions, we compare the resource
requirements of several implementations of the $1$-UpCCGSD and UCCGSD ans\"atze. The
benchmarks include naive ladder decompositions of Pauli rotations in the JW and BK mappings,
the Clifford-based synthesis method of Ref.~\cite{Pauli_compilation}, denoted below as GdBM,
and implementations based on the FSN and MSN constructions.

{\revised{20}{black}
All circuits are transpiled with Qiskit's \texttt{transpile} function to the
$\{\mathrm{CX},U_3\}$ basis using \texttt{optimization\_level=3}~\cite{qiskit2024} with default seed. We consider
two connectivity settings: all-to-all connectivity, for which no coupling map is imposed, and
a restricted $2\times N$ nearest-neighbor layout, with the coupling map shown in
Fig.~\ref{fig:device_connectivity} and the initial layout fixed by the logical row ordering of
the construction. The former setting is representative of trapped-ion processors, while the
latter models planar architectures such as superconducting, neutral-atom, and silicon spin-qubit
devices~\cite{strohm_ion-based_2024,huang_superconducting_2020,wintersperger_neutral_2023,Theory2xN,Experiment2xN}.
\endrevised{20}}

We report two main resource metrics: the transpiled circuit depth and the total number of
entangling $\mathrm{CX}$ gates. Here, depth is the Qiskit depth in the
$\{\mathrm{CX},U_3\}$ basis after imposing the chosen connectivity, so parallel single- and
two-qubit layers each contribute one unit. Since entangling gates typically dominate the error
budget, the total $\mathrm{CX}$ count is reported separately.

{\revised{20}{black}
In Figs.~\ref{fig:compare_kup} and~\ref{fig:compare_ucc}, the additional metric
``$\mathrm{CX}$ gates per Pauli string'' is defined as the total $\mathrm{CX}$ count divided by
the number of Pauli-strings corresponding to the $\Gamma_{i_1,\dots,i_k}$ terms in the ansatz.
\endrevised{20}}

\subsubsection{   \texorpdfstring{$2 \times N$}{2 x N}  architectures}

The proposed MSN is
 tailored to hardware
 with $2\times N$ connectivity. In contrast, the Yordanov decomposition for FSN cannot be directly transpiled to linear or two-dimensional architectures, as the first qubit interacts with all the others. To address this limitation, we consider a 19-$\mathrm{CX}$ modification of the Yordanov decomposition in which the first and second qubits are swapped, and each $\mathrm{CX}_{1,3}$ gate is realized through a sequence of four $\mathrm{CX}$ operations:
    \begin{equation}
    \mathrm{CX}_{1,3} = \mathrm{CX}_{1,2},\mathrm{CX}_{2,3},\mathrm{CX}_{1,2},\mathrm{CX}_{2,3}.
    \end{equation}

{\revised{21}{black}The modified Yordanov decomposition, depicted in Appendix Fig.~\ref{fig:linear_yordanov}, was used for the $2\times N$ connectivity setting. The corresponding resource scaling is reported in Fig.~\ref{fig:compare_kup} as the routed FSN $2\times N$ baseline.

This 19-\(\mathrm{CX}\) circuit is not claimed to be an optimal
restricted-connectivity implementation of an FSN double-excitation rotation. It is used only as a
transparent routed version of the standard Yordanov baseline. In particular, other
decompositions adapted directly to the \(2\times2\) geometry can have a smaller CX count.
Therefore, the reported \(2\times N\) improvements should be interpreted as improvements
over the specified routed FSN/Yordanov baseline, not as a comparison against all possible
FSN-based restricted-connectivity implementations.
\endrevised{21}
}

 \subsubsection{All-to-all connectivity architectures}

 The results in Fig.~\ref{fig:compare_kup} for $1$-UpCCGSD and in Fig.~\ref{fig:compare_ucc} for UCCGSD   demonstrate that SNs significantly outperform traditional approaches
  \DIFadd{that do} not use ancillas or mid-circuit
  \DIFadd{measurements} in terms of both depth and $\mathrm{CX}$ count. Among them, the proposed MSN constructions consistently yield the most compact circuits. In contrast, circuits based on ladder decompositions in the JW and BK mappings result in the highest circuit depths and gate counts among all tested methods.

The GdBM method provides substantial improvements over ladder-based approaches, but does not match the performance of SN-based alternatives \DIFadd{as the number of qubits increases}. However, the proposed MSN circuits yield an additional $\sim$55\% reduction in depth and $\sim$40\% reduction in entangling gate count relative to FSN-based circuits    \DIFadd{restricted} to $2\times N$ connectivity.

It should be noted that the standard Qiskit transpilation used in our simulation is unlikely to produce an optimal circuit under    \DIFadd{limited-connectivity} constraints. Therefore, in our noise-inclusive simulations, both ladder and GdBM decompositions were evaluated only under all-to-all connectivity.



 \subsection{Specific noise models}

In addition to these static resource metrics, we also examine the robustness of the synthesized
$1$-UpCCGSD circuits under realistic quantum noise. In particular, we simulate the behavior of
the corresponding VQE algorithm under various noise models, allowing us to assess the practical
viability of each construction beyond the ideal gate-based model.

We describe a noisy quantum channel $\Lambda$ acting on a state $\rho$ by Kraus operators
$\{E_i\}$:
  \begin{align}
        &\Lambda(\rho) = \sum_i E_i \rho E_i^\dagger,\\
        &\sum_i E_i^\dagger E_i = I.
    \end{align}
Each noisy gate is modeled as an ideal unitary $U$ followed by a noise channel
$\Lambda_{\mathrm{noise}}$:
    \begin{equation}
        \mathcal{E}_U(\rho) = \Lambda_{\mathrm{noise}}\!\left(U\rho U^\dagger\right).
    \end{equation}

We consider the following channels acting on a two-qubit subsystem $A$, since two-qubit
operations typically dominate the hardware error budget:
    \begin{itemize}
        \item two-qubit depolarizing channel:
        \begin{equation}
        \label{eq:D}
        \Lambda_D^A(\rho) = (1 - p)\rho + \frac{p}{4}\mathrm{Tr}_{A}(\rho) \otimes I_A,
        \end{equation}
        where $I_A$ is the identity on subsystem $A$.

        \item correlated two-qubit bit-flip channel:
        \begin{equation}
            \quad\quad\quad\Lambda_X^A(\rho) = (1 - p)\rho + p (X\otimes X)_A\rho (X\otimes X)_A.
        \end{equation}
        The correlated two-qubit $Y$- and $Z$-error channels are defined analogously.
    \end{itemize}
{\revised{22}{black}For the Pauli-channel benchmarks, the circuits were transpiled to the
$\{\mathrm{CX},U_3\}$ basis.\endrevised{22}}

\textit{Noise model for superconducting devices}. We use a phenomenological model based on
calibration data from the 156-qubit IBM device \texttt{ibm\_kingston}~\cite{ibm}. The reported
parameters are averaged over the device qubits. The circuits are transpiled to
{\revised{23}{black}the native basis $\{CZ,R_{zz},R_x,R_z\}$\endrevised{23}}.
The model combines thermal-relaxation errors parameterized by the measured $T_1$ and $T_2$ times,
with zero equilibrium excited-state population, and depolarizing channels calibrated to the
reported single- and two-qubit gate infidelities. Qiskit's thermal-relaxation channel incorporates
both energy relaxation and dephasing~\cite{super_noise}. The conversion from average gate
infidelity to depolarizing probability and the thermal-relaxation details are summarized in
Appendix~\ref{app:fid-noise}. The parameters used in the superconducting-device simulations are:
  \begin{equation}
        \label{eq:super_noise_params}
        \quad\quad\quad\begin{aligned}
            &T_1 = 264\:\mu\text{s},\\
            &T_2 = 162\:\mu\text{s},\\
            &t_1=t_{R_x}=t_{R_z}=0,\\
            &t_2=t_{CZ}=t_{R_{zz}} = 68\:\text{ns},\\
            &\varepsilon_1 = \varepsilon_{R_z} = \varepsilon_{R_x} = 4.4\cdot10^{-4},\\
            &\varepsilon_2 = \varepsilon_{CZ}=\varepsilon_{R_{zz}} = 5.5\cdot10^{-3}.
        \end{aligned}
    \end{equation} 

 \begingroup
\revised{23}{black}
In this superconducting model, both $R_x$ and $R_z$ are assigned zero duration for the thermal
relaxation part of the noise model, but they are not treated as noiseless: the averaged
single-qubit depolarizing error is still applied to both gates. Thermal relaxation and
depolarizing errors are attached after transpilation to the native gates $CZ$, $R_{zz}$, $R_x$,
and $R_z$. Relaxation errors on delay instructions due to transpilation are included. The connectivity restrictions enter through the circuit
family being compared, e.g., the routed $2\times N$ circuits or all-to-all baselines.
\endrevised{23}
\endgroup

\textit{Noise model for trapped-ion devices}. We construct an analogous phenomenological
model using calibration data from the IonQ Forte device. The circuits are transpiled to
{\revised{24}{black}the same $\{CZ,R_{zz},R_x,R_z\}$ basis.\endrevised{24}}
As in the superconducting model, thermal-relaxation errors specified by the reported $T_1$ and
$T_2$ times are combined with depolarizing channels calibrated to the reported single- and
two-qubit gate infidelities and timings~\cite{ionq}. The trapped-ion parameters used in this work
are:
  \begin{equation}
        \label{eq:ion_noise_params}
        \quad\quad\quad\begin{aligned}
            &T_1 = 188\:\text{s},\\
            &T_2 = 0.95\:\text{s},\\
            &t_1=t_{R_x}=t_{R_z} = 63\:\mu\text{s},\\
            &t_2=t_{CZ}=t_{R_{zz}} = 650\:\mu\text{s},\\
            &\varepsilon_1 = \varepsilon_{R_z} = \varepsilon_{R_x} = 2.0\cdot10^{-4},\\
            &\varepsilon_2 = \varepsilon_{CZ}=\varepsilon_{R_{zz}} = 6.2\cdot10^{-3}.
        \end{aligned}
    \end{equation} 
\revised{24}{black}
For the trapped-ion model, we use the same native basis and post-transpilation noise insertion as above, changing only the gate durations used in the relaxation channel: \(t_{R_x}=t_{R_z}=63\,\mu\mathrm{s}\) and \(t_{CZ}=t_{R_{zz}}=650\,\mu\mathrm{s}\).
\endrevised{24}
To explore different noise regimes, we introduce a noise multiplier $\lambda$ that scales
the depolarizing probabilities and the relaxation exponents as specified below.

\begin{figure}[tbp]
    \centering 
    \includegraphics[width=\linewidth]{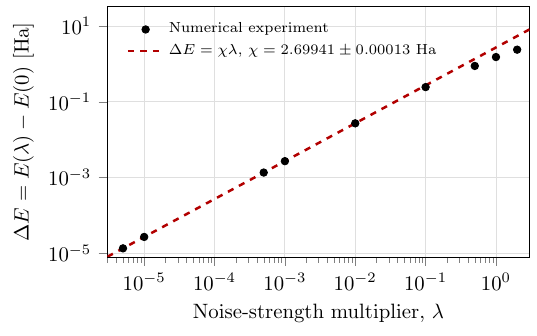}
    \caption{\label{fig:representative-susceptibility}\revised{25}{black}Representative extraction of the susceptibility $\chi$ for LiH on 10 qubits using the MSN $2\times N$ circuit under the superconducting noise model.
The panel shows the optimized energy shift $\Delta E_{\mathrm{opt}}(\lambda)=E_{\mathrm{opt}}(\lambda)-E_{\mathrm{opt}}(0)$ over the available multiplier sweep, together with the least-squares fit constrained through the origin to the four smallest nonzero values of $\lambda$. The quoted uncertainty is the standard error of the fitted slope.\endrevised{25}}
\end{figure}

{\revised{25}{black}
Next we define the noise susceptibility
\(\chi\) as the leading-order slope of the optimized VQE energy with respect to a scalar
noise-strength variable \(x\) to quantify circuit-level sensitivity to weak noise:
\begin{equation}
    \Delta E_{\mathrm{opt}}(x)
    \equiv E_{\mathrm{opt}}(x)-E_{\mathrm{opt}}(0)
    = \chi x + \mathcal{O}(x^2).
    \label{eq:noise_susceptibility}
\end{equation}
Here \(E_{\mathrm{opt}}(x)\) is the VQE energy after optimization of all variational parameters
at noise strength \(x\). The noiseless optimum is used only as the initial point for each noisy
optimization; at every nonzero \(x\), all parameters are reoptimized with the same L-BFGS-B
optimizer. We estimate \(\chi\) from an ordinary least-squares fit constrained through the
origin,
\(\Delta E_{\mathrm{opt}}(x)=\chi x\), using the four smallest nonzero noise values. The
uncertainty shown in Fig.~\ref{fig:representative-susceptibility} is the standard error of the
fitted slope obtained from the regression residuals. All noisy expectation values are evaluated
by exact density-matrix simulation. Throughout this section, \(x=p\) for the Pauli-channel
benchmarks and \(x=\lambda\) for the hardware-inspired benchmarks.
\endrevised{25}}

\subsubsection{Comparison across Pauli noise channels}

{
In Tables~\DIFadd{\ref{tab:h2_4q}--\ref{tab:lih_12q}}, boldface marks the lowest value in each column, while italics
mark the second-lowest distinct value. Ties are retained after rounding.

 }
  
\begin{table}[tbp]
     \centering
    \caption{\label{tab:h2_4q}\DIFaddFL{Pauli-channel noise susceptibility $\chi$ for H$_2$ on 4 qubits. }}
    \begin{adjustbox}{max width=\linewidth}
    \begin{tabular*}{\linewidth}{@{\extracolsep{\fill}} l c c c c c @{}}
    \toprule
    \textbf{Method} & \textbf{D} & \textbf{X} & \textbf{Y} & \textbf{Z} & \textbf{\# CX gates} \\
    \midrule
    JW & $49.1$ & $21.6$ & $40.2$ & $46.7$ & $49$ \\
    BK & $38.1$ & $34.3$ & $30.9$ & $31.0$ & $35$ \\
    JW GdBM & $24.2$ & $\mathit{16.4}$ & $23.2$ & $19.0$ & $21$ \\
    BK GdBM & $19.1$ & $19.6$ & $19.7$ & $\mathbf{0.7}$ & $\mathbf{15}$ \\
    FSN a-t-a & $\mathbf{14.9}$ & $17.2$ & $\mathit{15.1}$ & $\mathit{7.9}$ & $\mathbf{15}$ \\
    FSN $2\times N$ & $19.5$ & $21.1$ & $18.9$ & $10.8$ & $21$ \\
    \midrule
    MSN $2\times N$ & $\mathit{18.3}$ & $\mathbf{13.9}$ & $\mathbf{7.3}$ & $13.6$ & $\mathit{16}$ \\
    \bottomrule
    \end{tabular*}
    \end{adjustbox}
    \end{table}

\begin{table}[tbp]
     \centering
    \caption{\DIFaddFL{Pauli-channel noise susceptibility $\chi$ for H$_2$ on 8 qubits. }}
    \begin{adjustbox}{max width=\linewidth}
    \begin{tabular*}{\linewidth}{@{\extracolsep{\fill}} l c c c c c @{}}
    \toprule
    \textbf{Method} & \textbf{D} & \textbf{X} & \textbf{Y} & \textbf{Z} & \textbf{\# CX gates} \\
    \midrule
    JW & $715$ & $297$ & $612$ & $555$ & $417$ \\
    BK & $902$ & $567$ & $796$ & $735$ & $394$ \\
    JW GdBM & $487$ & $394$ & $452$ & $400$ & $151$ \\
    BK GdBM & $460$ & $369$ & $385$ & $278$ & $\mathit{132}$ \\
    FSN a-t-a & $\mathit{217}$ & $\mathit{255}$ & $\mathit{218}$ & $\mathbf{109}$ & $170$ \\
    FSN $2\times N$ & $267$ & $305$ & $254$ & $135$ & $134$ \\
    \midrule
    MSN $2\times N$ & $\mathbf{171}$ & $\mathbf{126}$ & $\mathbf{103}$ & $\mathit{124}$ & $\mathbf{96}$ \\
    \bottomrule
    \end{tabular*}
    \end{adjustbox}
    \end{table}

\begin{table}[tbp]
     \centering
    \caption{\DIFaddFL{Pauli-channel noise susceptibility $\chi$ for LiH on 8 qubits. }}
    \begin{adjustbox}{max width=\linewidth}
    \begin{tabular*}{\linewidth}{@{\extracolsep{\fill}} l c c c c c @{}}
    \toprule
    \textbf{Method} & \textbf{D} & \textbf{X} & \textbf{Y} & \textbf{Z} & \textbf{\# CX gates} \\
    \midrule
    JW & $634$ & $\mathit{220}$ & $531$ & $616$ & $417$ \\
    BK & $696$ & $409$ & $591$ & $610$ & $394$ \\
    JW GdBM & $539$ & $386$ & $477$ & $464$ & $151$ \\
    BK GdBM & $494$ & $369$ & $427$ & $309$ & $\mathit{132}$ \\
    FSN a-t-a & $\mathit{230}$ & $293$ & $\mathit{230}$ & $\mathbf{96}$ & $170$ \\
    FSN $2\times N$ & $282$ & $346$ & $269$ & $\mathit{125}$ & $134$ \\
    \midrule
    MSN $2\times N$ & $\mathbf{205}$ & $\mathbf{149}$ & $\mathbf{123}$ & $146$ & $\mathbf{96}$ \\
    \bottomrule
    \end{tabular*}
    \end{adjustbox}
    \end{table}

\begin{table}[tbp]
     \centering
    \caption{\DIFaddFL{Pauli-channel noise susceptibility $\chi$ for LiH on 10 qubits. }}
    \begin{adjustbox}{max width=\linewidth}
    \begin{tabular*}{\linewidth}{@{\extracolsep{\fill}} l c c c c c @{}}
    \toprule
    \textbf{Method} & \textbf{D} & \textbf{X} & \textbf{Y} & \textbf{Z} & \textbf{\# CX gates} \\
    \midrule
    JW & $1022$ & $\mathit{344}$ & $866$ & $917$ & $749$ \\
    BK & $1484$ & $825$ & $1269$ & $1184$ & $898$ \\
    JW GdBM & $1632$ & $1258$ & $1377$ & $1368$ & $434$ \\
    BK GdBM & $1773$ & $1462$ & $1578$ & $1162$ & $463$ \\
    FSN a-t-a & $\mathit{318}$ & $408$ & $\mathit{334}$ & $\mathbf{114}$ & $\mathit{216}$ \\
    FSN $2\times N$ & $393$ & $484$ & $387$ & $\mathit{155}$ & $276$ \\
    \midrule
    MSN $2\times N$ & $\mathbf{293}$ & $\mathbf{209}$ & $\mathbf{142}$ & $206$ & $\mathbf{160}$ \\
    \bottomrule
    \end{tabular*}
    \end{adjustbox}
    \end{table}

\begin{table}[tbp]
    \centering
    \caption{\label{tab:lih_12q}\DIFaddFL{Pauli-channel noise susceptibility $\chi$ for LiH on 12 qubits. }}
    \begin{adjustbox}{max width=\linewidth}
    \begin{tabular*}{\linewidth}{@{\extracolsep{\fill}} l c c c c c @{}}
    \toprule
    \DIFaddFL{\textbf{Method} }& \DIFaddFL{\textbf{D} }& \DIFaddFL{\textbf{X} }& \DIFaddFL{\textbf{Y} }& \DIFaddFL{\textbf{Z} }& \DIFaddFL{\textbf{\# CX gates} }\\
    \midrule
    \DIFaddFL{JW }& \DIFaddFL{$1698$ }& \DIFaddFL{$591$ }& \DIFaddFL{$1487$ }& \DIFaddFL{$1418$ }& \DIFaddFL{$1364$ }\\
    \DIFaddFL{BK }& \DIFaddFL{$2088$ }& \DIFaddFL{$1003$ }& \DIFaddFL{$1777$ }& \DIFaddFL{$1695$ }& \DIFaddFL{$1555$ }\\
    \DIFaddFL{JW GdBM }& \DIFaddFL{$1504$ }& \DIFaddFL{$1231$ }& \DIFaddFL{$1265$ }& \DIFaddFL{$1183$ }& \DIFaddFL{$403$ }\\
    \DIFaddFL{BK GdBM }& \DIFaddFL{$1669$ }& \DIFaddFL{$1206$ }& \DIFaddFL{$1415$ }& \DIFaddFL{$1250$ }& \DIFaddFL{$425$ }\\
    \DIFaddFL{FSN a-t-a }& \DIFaddFL{$\mathit{426}$ }& \DIFaddFL{$\mathit{554}$ }& \DIFaddFL{$\mathit{429}$ }& \DIFaddFL{$\mathbf{159}$ }& \DIFaddFL{$\mathit{321}$ }\\
    \DIFaddFL{FSN $2\times N$ }& \DIFaddFL{$525$ }& \DIFaddFL{$653$ }& \DIFaddFL{$502$ }& \DIFaddFL{$\mathit{212}$ }& \DIFaddFL{$411$ }\\
    \midrule
    \DIFaddFL{MSN $2\times N$ }& \DIFaddFL{$\mathbf{393}$ }& \DIFaddFL{$\mathbf{276}$ }& \DIFaddFL{$\mathbf{192}$ }& \DIFaddFL{$273$ }& \DIFaddFL{$\mathbf{240}$ }\\
    \bottomrule
    \end{tabular*}
    \end{adjustbox}
    \end{table}

The results in Tables~\DIFadd{\ref{tab:h2_4q}--\ref{tab:lih_12q}} show that the most noise-robust compilation depends on
the error channel. For H$_2$ on four qubits, MSN has the lowest susceptibilities under the
correlated $X$ and $Y$ channels, FSN a-t-a has the lowest depolarizing susceptibility, and BK
GdBM has the lowest susceptibility under correlated $Z$ noise. For all larger instances in
Tables~II--V, MSN has the lowest susceptibility under the depolarizing, $X$, and $Y$ channels,
whereas FSN a-t-a remains the least susceptible under the correlated $Z$ channel.

Although JW ladder circuits are generally among the most noise-sensitive constructions, their
susceptibility to the correlated $X$ channel is consistently lower than their susceptibility to
the other Pauli channels. This channel dependence reflects how errors propagate through a given
decomposition. For example, in the four-qubit BK GdBM circuit, correlated $Z$ errors leave the
Hartree--Fock reference state, corresponding to zero excitation angles, largely unaffected and
therefore have only a small effect on the energy near that reference state. A related mechanism
for JW ladder decompositions is discussed in Appendix~\ref{app:channel_dependent_noise}.

\subsubsection{Phenomenological model for superconducting and ion devices}

\begin{table}[tbp]
\centering
\caption{\label{tab:super}\revised{26}{black}Noise susceptibility $\chi$ for the superconducting hardware-inspired noise model.\endrevised{26}}
\begin{adjustbox}{max width=\linewidth}
\begin{tabular*}{\linewidth}{@{\extracolsep{\fill}} l c c c c c @{}}
\toprule
\textbf{Method} & H$_2$, 4q & H$_2$, 8q & LiH, 8q & LiH, 10q  & \DIFaddFL{LiH, 12q} \\
\midrule
JW & $0.37$ &    \DIFaddFL{$6.39$}& $5.33$ &    \DIFaddFL{$9.08$ }& \DIFaddFL{$14.52$}\\
BK & $0.32$ & $7.76$ & $5.82$ &    \DIFaddFL{$13.26$ }& \DIFaddFL{$18.17$}\\
JW GdBM & $0.20$ & $4.64$ &    \DIFaddFL{$3.99$}&    \DIFaddFL{$7.61$ }& \DIFaddFL{$13.20$}\\
BK GdBM &    \DIFaddFL{$\mathit{0.15}$}& $4.41$ & $4.08$ &    \DIFaddFL{$8.73$ }& \DIFaddFL{$14.86$ }\\
\DIFaddFL{FSN a-t-a }& \DIFaddFL{$\mathbf{0.13}$ }& \DIFaddFL{$\mathit{1.85}$ }& \DIFaddFL{$\mathit{1.96}$ }& \DIFaddFL{$\mathit{2.72}$ }& \DIFaddFL{$\mathit{3.66}$}\\
FSN $2\times N$ & $0.16$ & $2.20$ & $2.33$ & $3.24$  & \DIFaddFL{$4.35$}\\
\midrule
MSN $2\times N$ & $0.17$ & $\mathbf{1.69}$ & $\mathbf{1.88}$ & $\mathbf{2.70}$  & \DIFaddFL{$\mathbf{3.62}$}\\
\bottomrule
\end{tabular*}
\end{adjustbox}
\end{table}

\begin{table}[tbp]
\centering
\caption{\label{tab:ion}\revised{26}{black}Noise susceptibility $\chi$ for the trapped-ion hardware-inspired noise model.\endrevised{26}}
\begin{adjustbox}{max width=\linewidth}
\begin{tabular*}{\linewidth}{@{\extracolsep{\fill}} l c c c c c @{}}
\toprule
\textbf{Method} & H$_2$, 4q & H$_2$, 8q & LiH, 8q & LiH, 10q  & \DIFaddFL{LiH, 12q}\\
\midrule
JW &    \DIFaddFL{$0.37$}&    \DIFaddFL{$6.60$}&    \DIFaddFL{$5.52$}&    \DIFaddFL{$9.60$ }& \DIFaddFL{$15.59$}\\
BK &    \DIFaddFL{$0.31$}&    \DIFaddFL{$8.09$}&    \DIFaddFL{$6.02$}&    \DIFaddFL{$14.14$ }& \DIFaddFL{$19.49$}\\
JW GdBM &    \DIFaddFL{$0.20$}&    \DIFaddFL{$4.79$}&    \DIFaddFL{$3.96$}&    \DIFaddFL{$7.79$ }& \DIFaddFL{$13.64$}\\
BK GdBM & $\mathit{0.14}$ &    \DIFaddFL{$4.38$}&    \DIFaddFL{$4.04$}&    \DIFaddFL{$8.91$ }& \DIFaddFL{$15.45$}\\
FSN a-t-a & $\mathbf{0.12}$ &    \DIFaddFL{$\mathit{1.75}$}&    \DIFaddFL{$\mathit{1.83}$}&    \DIFaddFL{$\mathit{2.53}$ }& \DIFaddFL{$\mathit{3.38}$ }\\
\DIFaddFL{FSN $2\times N$ }& \DIFaddFL{$0.15$ }& \DIFaddFL{$2.10$ }& \DIFaddFL{$2.20$ }& \DIFaddFL{$3.05$ }& \DIFaddFL{$4.07$}\\
\midrule
MSN    \DIFaddFL{$2\times N$}& $0.15$ &    \DIFaddFL{$\mathbf{1.55}$}&    \DIFaddFL{$\mathbf{1.71}$}&    \DIFaddFL{$\mathbf{2.47}$ }& \DIFaddFL{$\mathbf{3.31}$}\\
\bottomrule
\end{tabular*}
\end{adjustbox}
\end{table}

{\revised{27}{black}For the hardware-inspired noise analysis, the susceptibility is the derivative with respect to the multiplier $\lambda$:
\begin{equation}
\chi = \left.\frac{d E_{\mathrm{opt}}}{d\lambda}\right|_{\lambda = 0}.
\end{equation}
\ 

Before noise insertion, the depolarizing probabilities are scaled as
$p_g(\lambda)=\lambda p_g^{(0)}$, while the relaxation exponents are scaled as
$t_g/T_{1,2}\mapsto \lambda t_g/T_{1,2}$, including explicit delays. Thus $\lambda=1$
corresponds to the calibrated reference model and $\lambda=0$ to the noiseless circuit. 
\endrevised{27}}

{\revised{28}{black}Tables~\ref{tab:super} and~\ref{tab:ion} show that the swap-network circuits have the lowest or near-lowest susceptibility in most hardware-inspired instances, while the GdBM variants generally outperform the corresponding JW and BK ladder circuits. The superconducting and trapped-ion models give the same qualitative ordering. For FSN, the superconducting columns use the $2\times N$ circuit and the trapped-ion columns use the all-to-all circuit.\endrevised{28}}

\subsection{Error sources in VQE}

{\revised{29}{black}Tables~\ref{tab:h2_8q}--\ref{tab:lih_12q} decompose the hardware-inspired susceptibility into contributions from the full noise model, combined one- and two-qubit depolarizing noise, and two-qubit depolarizing noise alone.\endrevised{29}}

\begin{table}[H]
\centering
\caption{\revised{29}{black}Hardware-inspired noise susceptibility $\chi$ for H$_2$ on 8 qubits.\endrevised{29}}
\label{tab:h2_8q}
\begin{adjustbox}{max width=\linewidth}
\begin{tabular*}{\linewidth}{@{\extracolsep{\fill}} l c c c | c c c@{}}
\toprule
\textbf{Method} & \textbf{sc Full} & \textbf{1q}\&\textbf{2q} & \textbf{2q} & \textbf{ion Full} & \textbf{1q} \& \textbf{2q} & \textbf{2q} \\
\midrule
JW &    \DIFaddFL{$6.39$}& $5.65$ & $4.41$ & $6.60$ &    \DIFaddFL{$5.57$}& $4.95$ \\
BK & $7.76$ & $6.70$ & $5.25$ &    \DIFaddFL{$8.09$}&    \DIFaddFL{$6.49$}& $5.90$ \\
JW GdBM & $4.64$ &    \DIFaddFL{$4.04$}& $3.01$ &    \DIFaddFL{$4.79$}&    \DIFaddFL{$3.95$}& $3.38$ \\
BK GdBM & $4.41$ & $3.90$ & $2.94$ &    \DIFaddFL{$4.38$}&    \DIFaddFL{$3.75$}& $3.30$ \\
FSN  &    \DIFaddFL{$\mathit{1.85}$}&    \DIFaddFL{$\mathit{1.76}$}&    \DIFaddFL{$\mathit{1.22}$}&    \DIFaddFL{$\mathit{1.75}$}&    \DIFaddFL{$\mathit{2.02}$}& $\mathit{1.37}$ \\
\midrule
MSN  & $\mathbf{1.69}$ & $\mathbf{1.59}$ & $\mathbf{1.02}$ &    \DIFaddFL{$\mathbf{1.55}$}&    \DIFaddFL{$\mathbf{1.39}$}& $\mathbf{1.14}$ \\
\bottomrule
\end{tabular*}
\end{adjustbox}
\end{table}

\begin{table}[H]
\centering
\caption{\revised{29}{black}Hardware-inspired noise susceptibility $\chi$ for LiH on 8 qubits.\endrevised{29}}
\label{tab:lih_8q}
\begin{adjustbox}{max width=\linewidth}
\begin{tabular*}{\linewidth}{@{\extracolsep{\fill}} l c c c | c c c@{}}
\toprule
\textbf{Method} & \textbf{sc Full} & \textbf{1q}\&\textbf{2q} & \textbf{2q} & \textbf{ion Full} & \textbf{1q}\&\textbf{2q} & \textbf{2q} \\
\midrule
JW & $5.33$ & $4.60$ & $3.53$ &    \DIFaddFL{$5.52$}& $4.45$ & $3.97$ \\
BK & $5.82$ & $5.00$ & $3.86$ &    \DIFaddFL{$6.02$}& $4.85$ & $4.34$ \\
JW GdBM &    \DIFaddFL{$3.99$}& $3.60$ & $2.61$ &    \DIFaddFL{$3.96$}& $3.38$ & $2.93$ \\
BK GdBM & $4.08$ & $3.71$ & $2.73$ &    \DIFaddFL{$4.04$}& $3.51$ & $3.07$ \\
FSN &    \DIFaddFL{$\mathit{1.96}$}&    \DIFaddFL{$\mathit{1.84}$}&    \DIFaddFL{$\mathit{1.28}$}&    \DIFaddFL{$\mathit{1.83}$}& $\mathit{1.69}$ & $\mathit{1.43}$ \\
\midrule
MSN  & $\mathbf{1.88}$ & $\mathbf{1.77}$ & $\mathbf{1.13}$ &    \DIFaddFL{$\mathbf{1.71}$}& $\mathbf{1.56}$ & $\mathbf{1.27}$ \\
\bottomrule
\end{tabular*}
\end{adjustbox}
\end{table}

\begin{table}[H]
\centering
\caption{\revised{29}{black}Hardware-inspired noise susceptibility $\chi$ for LiH on 10 qubits.\endrevised{29}}
\label{tab:lih_10q}
\begin{adjustbox}{max width=\linewidth}
\begin{tabular*}{\linewidth}{@{\extracolsep{\fill}} l c c c | c c c@{}}
\toprule
\textbf{Method} & \textbf{sc Full} & \textbf{1q}\&\textbf{2q} & \textbf{2q} & \textbf{ion Full} & \textbf{1q}\&\textbf{2q} & \textbf{2q} \\
\midrule
JW &    \DIFaddFL{$9.08$}& $7.62$ &    \DIFaddFL{$5.91$}&    \DIFaddFL{$9.60$}&    \DIFaddFL{$7.41$}& $6.65$ \\
BK &    \DIFaddFL{$13.26$}&    \DIFaddFL{$10.99$}&    \DIFaddFL{$8.75$}&    \DIFaddFL{$14.14$}&    \DIFaddFL{$10.84$}&    \DIFaddFL{$9.83$}\\
JW GdBM &    \DIFaddFL{$7.61$}& $6.62$ & $4.96$ &    \DIFaddFL{$7.79$}& $6.32$ & $5.57$ \\
BK GdBM &    \DIFaddFL{$8.73$}& $7.53$ & $5.83$ &    \DIFaddFL{$8.91$}& $7.32$ & $6.55$ \\
FSN &    \DIFaddFL{$\mathit{2.72}$}&    \DIFaddFL{$\mathit{2.57}$}&    \DIFaddFL{$\mathit{1.77}$ }&    \DIFaddFL{$\mathit{2.53}$}& $\mathit{2.34}$ &    \DIFaddFL{$\mathit{1.98}$ }\\
\midrule
MSN  & $\mathbf{2.70}$ & $\mathbf{2.55}$ & $\mathbf{1.63}$ &    \DIFaddFL{$\mathbf{2.47}$}& $\mathbf{2.24}$ & $\mathbf{1.83}$ \\
\bottomrule
\end{tabular*}
\end{adjustbox}
\end{table}

\begin{table}[H]
\centering
\caption{\revised{29}{black}Hardware-inspired noise susceptibility $\chi$ for LiH on 12 qubits.\endrevised{29}}
\label{tab:lih_12q}
\setlength{\tabcolsep}{1pt}
\begin{adjustbox}{max width=\linewidth}
\begin{tabular*}{\linewidth}{@{\extracolsep{\fill}} l c c c | c c c@{}}
\toprule
\DIFaddFL{\textbf{Method} }& \DIFaddFL{\textbf{sc Full} }& \DIFaddFL{\textbf{1q\&2q} }& \DIFaddFL{\textbf{2q} }& \DIFaddFL{\textbf{ion Full} }& \DIFaddFL{\textbf{1q\&2q} }& \DIFaddFL{\textbf{2q} }\\
\midrule
\DIFaddFL{JW }& \DIFaddFL{$14.52$ }& \DIFaddFL{$11.88$ }& \DIFaddFL{$9.33$ }& \DIFaddFL{$15.59$ }& \DIFaddFL{$11.63$ }& \DIFaddFL{$10.48$ }\\
\DIFaddFL{BK }& \DIFaddFL{$18.17$ }& \DIFaddFL{$14.62$ }& \DIFaddFL{$11.57$ }& \DIFaddFL{$19.49$ }& \DIFaddFL{$14.37$ }& \DIFaddFL{$12.99$ }\\
\DIFaddFL{JW GdBM }& \DIFaddFL{$13.20$ }& \DIFaddFL{$11.23$ }& \DIFaddFL{$8.59$ }& \DIFaddFL{$13.64$ }& \DIFaddFL{$10.84$ }& \DIFaddFL{$9.65$ }\\
\DIFaddFL{BK GdBM }& \DIFaddFL{$14.86$ }& \DIFaddFL{$12.52$ }& \DIFaddFL{$9.65$ }& \DIFaddFL{$15.45$ }& \DIFaddFL{$12.13$ }& \DIFaddFL{$10.84$ }\\
\DIFaddFL{FSN }& \DIFaddFL{$\mathit{3.66}$ }& \DIFaddFL{$\mathit{3.45}$ }& \DIFaddFL{$\mathit{2.37}$ }& \DIFaddFL{$\mathit{3.38}$ }& \DIFaddFL{$\mathit{3.15}$ }& \DIFaddFL{$\mathit{2.67}$ }\\
\midrule
\DIFaddFL{MSN  }& \DIFaddFL{$\mathbf{3.62}$ }& \DIFaddFL{$\mathbf{3.43}$ }& \DIFaddFL{$\mathbf{2.19}$ }& \DIFaddFL{$\mathbf{3.31}$ }& \DIFaddFL{$\mathbf{3.02}$ }& \DIFaddFL{$\mathbf{2.46}$ }\\
\bottomrule
\end{tabular*}
\end{adjustbox}
\end{table}

{\revised{29}{black}
Results for the susceptibility \(\chi\) are reported for two phenomenological hardware models---superconducting (left
columns, sc) and trapped-ion (right columns, ion)---and for three noise partitions:
\textbf{Full} (complete model),
\textbf{1q}$\&$\textbf{2q} (single- and two-qubit depolarization only; relaxation channels omitted),
and \textbf{2q} (two-qubit depolarization only). Boldface marks the minimum \(\chi\) in each column. For FSN, the superconducting columns use the $2\times N$ circuit and the trapped-ion columns use the all-to-all circuit.
\endrevised{29}}

\section{Conclusion}
\label{sec:conclusion}

\DIFadd{We introduced \textit{Majorana Swap Networks}, an MSWAP-based architecture that permutes Majorana operators and reduces parameterized circuit depth and two-qubit gate count for chemically motivated variational ansatzes.}

{\revised{30}{black}For UCCGSD on all-to-all hardware,\endrevised{30}} we developed a
cyclic-permutation strategy that {
  \revised{30}{black}lowers the auxiliary routing overhead required
to localize the $\mathcal{O}(M^4)$ possible double excitations to $\mathcal{O}(M^3)$ MSWAP
transpositions, where $M$ is the number of fermionic modes. For the complete UCCGSD ansatz,
applying all localized double-excitation rotations still gives an $\mathcal{O}(M^4)$ rotation
cost.\endrevised{30}} 
 For $k$-UpCCGSD on $2\times N$ connectivity, we introduced a more streamlined swap network and incorporated
excitation operators directly into the swap layers. Relative to the FSN baselines considered
here, this yields constant-factor
 reductions of up to $55$\% in depth and $40$\% in two-qubit
operations.

Numerical modeling
  \DIFadd{indicates} that these structural advantages
  \DIFadd{are often accompanied by} improved noise resilience    \DIFadd{on the tested small molecular instances.
}{\revised{31}{black}In the superconducting and trapped-ion noise models studied here, MSN circuits generally
exhibited the lowest or near-lowest energy susceptibility $\chi$. For individual Pauli error channels, however, the ranking
can depend on the decomposition and on how errors propagate through the circuit. This indicates
that robustness depends not only on total gate count, but also on the specific excitation
decomposition and error channel.\endrevised{31}}

Since the initial completion of this manuscript, more general fermionic-routing frameworks have
appeared. Ref.~\cite{maskara2025reconfigurable} achieves generic fermionic routing with
worst-case depth overhead $\mathcal{O}(\log N)$ by using $\mathcal{O}(1)$-depth multi-qubit
interleave swaps together with ancillas, mid-circuit measurement, and feedforward.
Ref.~\cite{constantinides2025lowdepth} implements arbitrary fermionic permutations in depth
$\mathcal{O}(\log^2 N)$ under all-to-all connectivity without ancillas or measurements. These
results strengthen the asymptotic theory of general fermionic routing and therefore go beyond
the fixed-baseline comparisons considered in this work. A straightforward black-box substitution
of such generic routing primitives into a $k$-UpCCGSD compilation would still yield an
$\mathcal{O}(N\log N)$-depth implementation of a layer, since the polylogarithmic routing
overhead would be incurred across $\mathcal{O}(N)$ structured excitation blocks. By contrast,
the FSN/MSN constructions studied here are tailored to the excitation structure of
$k$-UpCCGSD/UCCGSD and derive much of their practical benefit from constant-depth swap layers.
In this sense, the later routing results are best viewed as complementary to the present
approach: they improve the asymptotics of fully general fermionic routing, whereas our method
focuses on structured variational circuits, where ansatz-specific ordering can further reduce
circuit depth and gate count.

\section*{Code availability}

The experimental code and numerical data used for circuit generation,
resource estimates, and noise simulations are publicly available at
\url{https://github.com/DanilkaFish/SwapNetworks}. The repository
contains the circuit-generation code, utility scripts, numerical output
files, and a dependency specification for running the numerical
workflow.

\begin{acknowledgments}
S.A.F. acknowledges the support from the Foundation for the Advancement of Theoretical Physics
and Mathematics (BASIS) (Project № 23-2-10-15-1) and the Scholarships of the President of the
Russian Federation for postgraduate students.
\end{acknowledgments}



\medskip
\bibliography{main}



\onecolumngrid\newpage

\appendix
\setcounter{figure}{0}
\renewcommand\thefigure{\Alph{section}\arabic{figure}}
 \renewcommand\theHfigure{\Alph{section}.\arabic{figure}}

\newcommand{\DrawNodes}[3][4.8mm]{%

\begin{figure}[hbtp!]
    \centering
    \begin{tikzpicture}
        \begin{scope}[transform shape, thick, font=\fontsize{12}{12}\selectfont]
            \foreach \content [count=\i from 1] in {#2} {%
              \ifnum\i=1
                \node[] (\i) {\content};
              \else
                \node[right=#1 of \the\numexpr\i-1\relax] (\i) {\content};
              \fi
          }%
        \newdimen\BracketH
          \BracketH=10pt\relax
          \foreach \L/\R in {#3} {%
            \TopSquareBracket[black]{\L}{\R}{2pt}{\the\BracketH}%
            \advance\BracketH by -3pt\relax
        }%
          \end{scope}%
    \end{tikzpicture}
\end{figure}

}

\section{Proof of correctness for the cyclic permutation algorithm}
 \label{app:cyclic_proof}
 \label{sec:cycl_appendix}

  \begingroup
\revised{33}{black}

In this appendix we prove that the cyclic permutation algorithm described in
Sec.~\ref{sec:cyclic_permut} encounters every four-mode set that can appear in a
double excitation. More precisely, we show that every such set can be written,
at some step of the algorithm, as the union of two local fermionic pairs. By a
local fermionic pair we mean a pair of fermionic modes whose product
has two-qubit support in the modified JW ordering used in
Sec.~\ref{sec:cyclic_permut}. Therefore, when a four-mode set is encountered as
the union of two local pairs, the corresponding double-excitation operator has a
compact four-qubit representation.

Consider a stage of the algorithm in which the active set contains \(2K\)
fermionic modes. We write the current ordering of these modes as a list:
\begin{equation}
    L=(\ell_0,\ell_1,\ldots,\ell_{2K-1}).
\end{equation}
The local fermionic pairs are defined by reflection about the midpoint of the
list, as in Fig.~\ref{fig:cycl}(b):
\begin{equation}
    P_i(L)=\{\ell_i,\ell_{2K-1-i}\},
    \qquad i=0,\ldots,K-1.
    \label{eq:local_pairs_appendix}
\end{equation}
A four-mode set \(Q\) is exposed by the ordering \(L\) if it is the union of two
distinct local pairs:
\begin{equation}
    Q=P_i(L)\cup P_j(L),
    \qquad i\neq j.
    \label{eq:exposed_quartet_appendix}
\end{equation}
In this case, the qubit representation of the corresponding double-excitation
operator is supported on a compact four-qubit block in the current encoding.

We first analyze one sweep of a selected rotating mode. Let the initial active
list be
\begin{equation}
    L^0=(a,b_0,b_1,\ldots,b_{R-1}),
    \qquad R=2K-1,
    \label{eq:initial_list_appendix}
\end{equation}
where \(a\) is the rotating fermionic mode and \(b_0,\ldots,b_{R-1}\) are the
remaining active modes. All indices of the \(b\)'s are taken modulo \(R\). Let
\(\tau_r\) denote the adjacent transposition that swaps positions \(r\) and
\(r+1\), with the position index taken modulo \(2K\). We define
\begin{equation}
    L^t=\tau_{t-1}\tau_{t-2}\cdots\tau_0(L^0),
    \qquad \tau_{r+2K}\equiv \tau_r.
    \label{eq:Lt_definition_appendix}
\end{equation}
If \(t=2Kp+q\), where \(0\le q<2K\), then
\begin{equation}
\begin{aligned}
    L^t
    =
    (&b_p,b_{p+1},\ldots,b_{p+q-1},
    a,
    b_{p+q},\ldots,b_{p-1}).
\end{aligned}
\label{eq:list_after_t_appendix}
\end{equation}
Thus \(L^t\) describes the intermediate orderings obtained while the rotating
mode \(a\) is moved through the remaining \(2K-1\) modes.

The sequence has a half-period symmetry. Let
\begin{equation}
    H_K=K(2K-1).
\end{equation}
The list \(L^{t+H_K}\) is obtained from \(L^t\) by shifting all entries by
\(K\) positions. This shift maps the reflected pairs in
Eq.~\eqref{eq:local_pairs_appendix} to reflected pairs again, and therefore
\begin{equation}
    \{P_i(L^{t+H_K})\}_{i=0}^{K-1}
    =
    \{P_i(L^t)\}_{i=0}^{K-1}.
    \label{eq:half_period_pairing_appendix}
\end{equation}
Hence the second half of the full period produces no new local pairings. It is
therefore sufficient to consider a sweep of length \(H_K\).

\begin{lemma}
\label{lem:rotating_sweep}
During one sweep of the rotating mode \(a\), every four-mode set containing
\(a\) is exposed at least once.
\end{lemma}
\begin{proof}
For clarity, we first consider the full sequence obtained by allowing
\(p\) in Eq.~\eqref{eq:list_after_t_appendix} to run over all residues modulo
\(R=2K-1\). At the end of the proof we use the half-period symmetry
Eq.~\eqref{eq:half_period_pairing_appendix} to reduce this to
\(H_K=K(2K-1)\) transpositions.

For fixed \(p\) and \(d=0,\ldots,K-2\), define
\begin{equation}
    B_{p,d}=\{b_{p+d},b_{p-d-1}\}.
    \label{eq:Bpd_appendix}
\end{equation}
The integer \(d\) measures how far the two elements of the pair are separated in
the ordered list of \(b\)-modes. We prove by induction on \(d\) the following
statement:
\begin{equation}
    \mathcal{S}_d:\quad
    \text{for every } p\text{ and every } x\notin B_{p,d},
    \text{ the set }\{a,x\}\cup B_{p,d}\text{ is exposed.}
\end{equation}
Since every unordered pair of distinct \(b\)-modes can be written as
\(B_{p,d}\) for a suitable \(p\) and \(d\), this statement implies that every
four-mode set containing \(a\) is exposed.

First note the direct exposure mechanism. In the list \(L^t\) with
\(t=2Kp+q\), the two modes in \(B_{p,d}\) occupy reflected positions whenever
\begin{equation}
    d+1\le q\le 2K-d-2.
    \label{eq:Bpd_range_appendix}
\end{equation}
Thus \(B_{p,d}\) is one of the local fermionic pairs throughout this range of
\(q\). During the same range, the rotating mode \(a\) is paired with every mode
in the complementary part of the ordered list,
\begin{equation}
    b_{p+d+1}, b_{p+d+2},\ldots,b_{p-d-2},
    \label{eq:paired_with_a_appendix}
\end{equation}
where the indices are understood modulo \(R\). Therefore all four-mode sets
\begin{equation}
    \label{eq:range}
    \{a,x\}\cup B_{p,d},
    \qquad
    x\in\{b_{p+d+1},\ldots,b_{p-d-2}\},
\end{equation}
are exposed directly.

We now prove \(\mathcal{S}_d\) by induction. For \(d=0\), we have
\begin{equation}
    B_{p,0}=\{b_p,b_{p-1}\}.
\end{equation}
In this case the list in Eq.~\eqref{eq:paired_with_a_appendix} contains all
\(b\)-modes except \(b_p\) and \(b_{p-1}\). Hence every set
\(\{a,x\}\cup B_{p,0}\) with \(x\notin B_{p,0}\) is exposed directly, proving
\(\mathcal{S}_0\).

Assume now that \(\mathcal{S}_{d'}\) holds for all \(d'<d\), and consider the
pair
\begin{equation}
    B_{p,d}=\{b_{p+d},b_{p-d-1}\}.
\end{equation}
By the direct exposure argument above, all choices of \(x\) inside the interval
between these two endpoints are already covered by Eq.~\eqref{eq:range}. It
remains to consider the modes lying outside them, which can be written as
\begin{equation}
    x=b_{p+d-r},
    \qquad r=1,\ldots,2d.
    \label{eq:x_inside_interval_appendix}
\end{equation}
If \(r\) is odd, then the pair \(\{x,b_{p+d}\}\) has the form
\(B_{p',d'}\) with
\begin{equation}
    d'=\frac{r-1}{2}<d.
\end{equation}
By the induction hypothesis \(\mathcal{S}_{d'}\), the four-mode set
\begin{equation}
    \{a,x,b_{p+d},b_{p-d-1}\}
\end{equation}
is exposed. If \(r\) is even, then instead the pair
\(\{x,b_{p-d-1}\}\) has the form \(B_{p'',d''}\) with
\begin{equation}
    d''=\frac{2d-r}{2}<d.
\end{equation}
Again, the induction hypothesis implies that
\begin{equation}
    \{a,x,b_{p+d},b_{p-d-1}\}
\end{equation}
is exposed. Thus every remaining choice of \(x\) is covered, and
\(\mathcal{S}_d\) follows.

By induction, \(\mathcal{S}_d\) holds for all \(d=0,\ldots,K-2\). Hence the full
sequence exposes every four-mode set containing the rotating mode \(a\). Finally,
Eq.~\eqref{eq:half_period_pairing_appendix} shows that the second half of the
full sequence repeats the same local pairings as the first half. Since exposure
depends only on the set of local fermionic pairs, a sweep of
\(H_K=K(2K-1)\) transpositions is sufficient.
\end{proof}

\begin{lemma}
\label{lem:two_sweep_stage}
At a stage with \(2K\) active modes, suppose the algorithm performs one sweep
for a rotating mode \(a\) and one sweep for a second rotating mode \(b\). Then
this stage exposes every four-mode set in the active set that contains \(a\) or
\(b\).
\end{lemma}

\begin{proof}
By Lemma~\ref{lem:rotating_sweep}, the sweep of \(a\) exposes every four-mode
set containing \(a\). The same argument applies to the sweep of \(b\). Reversing
the direction of a sweep only changes the order in which the same reflected
pairings are encountered. Hence the two sweeps expose all four-mode sets that
contain at least one of the two rotating modes.
\end{proof}

\begin{theorem}
\label{thm:full_cyclic_coverage}
The full alternating cyclic permutation algorithm exposes every four-mode set of
fermionic modes.
\end{theorem}

\begin{proof}
Let
\begin{equation}
    Q=\{q_1,q_2,q_3,q_4\}
\end{equation}
be an arbitrary four-mode set. The algorithm proceeds in stages. At each stage,
two rotating modes are selected, swept through the current active set, and then
removed from the active set for the purpose of subsequent stages.

Let \(r\) be the first stage at which one of the four modes in \(Q\) is selected
as a rotating mode. Before stage \(r\), none of the modes in \(Q\) has been
removed, so all four modes of \(Q\) are still present in the active set at the
beginning of that stage. Since \(Q\) contains one of the two rotating modes
selected at stage \(r\), Lemma~\ref{lem:two_sweep_stage} implies that \(Q\) is
exposed during that stage as the union of two local fermionic pairs. Because
\(Q\) was arbitrary, the algorithm exposes every four-mode set.
\end{proof}

This completes the proof that the alternating cyclic procedure exposes every
four-mode set as the union of two local fermionic pairs at some step of the
algorithm. The transposition count and its scaling interpretation are discussed
in Sec.~\ref{sec:cyclic_permut}.

\endrevised{33}
\endgroup

 \section{Fidelity and \DIFadd{parameters of noise channels}}
\label{app:fid-noise}

\subsection{Depolarization}
\label{app:depol}

On a $d$-dimensional Hilbert space, the depolarizing channel with parameter $p \in [0,1]$ is
\begin{equation}
\Lambda_{\text{d}}(\rho)=(1-p)\,\rho + p\,\frac{\mathbb{I}_d}{d}.
\end{equation}

The \emph{average gate fidelity} of $\Lambda_{\text{d}}$ to the identity is
\begin{equation}
F_{\mathrm{avg}}(\Lambda_{\text{d}},\mathbb{I})
= \int \!\! d\psi \; \langle \psi | \Lambda_{\text{d}}(|\psi\rangle\!\langle\psi|) | \psi \rangle
= (1-p) + p\,\frac{1}{d}
= 1 - p\,\frac{d-1}{d}.
\label{eq:Favg-depol}
\end{equation}
Hence the \emph{average gate infidelity} $\varepsilon:=1-F_{\mathrm{avg}}$ and $p$ are related by
  \begin{equation}
\varepsilon = p\,\frac{d-1}{d}.
\label{eq:r-to-p}
\end{equation} 
Two common special cases are:
\begin{align}
d=2\ \text{(1q):}\quad & F_{\mathrm{avg}} = 1-\tfrac{p}{2}, \qquad p = 2\varepsilon, \\
d=4\ \text{(2q):}\quad & F_{\mathrm{avg}} = 1-\tfrac{3p}{4}, \qquad p = \tfrac{4}{3}\varepsilon.
\end{align}

\subsection{Transverse relaxation}
\label{app:transverse}

The reduced dynamics over a gate of duration \(\tau\) are well described on the Bloch sphere by
an affine map \cite{transv_relax}
\begin{equation}
\begin{aligned}
x' &= e^{-\tau/T_2}\,x,\\
y' &= e^{-\tau/T_2}\,y,\\
z' &= 1 + \bigl(z - 1\bigr)\,e^{-\tau/T_1},
\end{aligned}
\label{eq:bloch-transverse}
\end{equation}
where \(T_1\) is the energy-relaxation time and \(T_2\) the transverse (coherence) time. The
times satisfy the physical constraint
\begin{equation}
\frac{1}{T_2} = \frac{1}{2T_1} + \frac{1}{T_{\varphi}}, \qquad T_2 \le 2T_1,
\label{eq:T2-bound}
\end{equation}
with \(T_{\varphi}\) the \emph{pure dephasing} time.

For zero equilibrium excited-state population, a convenient Kraus representation of
Eq.~\eqref{eq:bloch-transverse} is obtained by composing an \emph{amplitude-damping} channel
with a \emph{pure-dephasing} channel, which supplies the additional phase decay required to
match the measured \(T_2\):
\begin{equation}
\Lambda_{\text{th}}(\tau) \;=\; \underbrace{\Lambda_{\text{P}}\!\bigl(\eta\bigr)}_{\text{pure dephasing}}
\;\circ\;
\underbrace{\Lambda_{\text{A}}\!\bigl(\gamma\bigr)}_{\text{amplitude damping}},
\label{eq:transverse_channel}
\end{equation}
with
\begin{equation}
\gamma = 1-e^{-\tau/T_1}, \qquad \eta = e^{-\tau/T_{\varphi}}, \qquad \frac{1}{T_{\varphi}} = \frac{1}{T_2} - \frac{1}{2T_1}.
\end{equation}

One convenient choice of Kraus operators for the amplitude-damping channel is
\DIFadd{
\begin{equation}
A_0=\begin{pmatrix}1&0\\[2pt]0&\sqrt{1-\gamma}\end{pmatrix},\qquad
A_1=\begin{pmatrix}0&\sqrt{\gamma}\\[2pt]0&0\end{pmatrix}.
\label{eq:gad-kraus}
\end{equation}
}
A two-element Kraus representation for pure dephasing is
\begin{equation}
P_0=\sqrt{\tfrac{1+\eta}{2}}\,\mathbb{I},\qquad
P_1=\sqrt{\tfrac{1-\eta}{2}}\,\sigma_z.
\end{equation}

To prove the decomposition Eq.~\eqref{eq:transverse_channel}, note that the zero-temperature
amplitude-damping channel with \(\gamma = 1 - e^{-\tau/T_1}\) acts on the Bloch vector as
\begin{equation}
\begin{aligned}
x' &= \sqrt{1-\gamma}\,x \;=\; e^{-\tau/(2T_1)}\,x,\\
y' &= \sqrt{1-\gamma}\,y \;=\; e^{-\tau/(2T_1)}\,y,\\
z' &= (1-\gamma)\,z + \gamma \;=\; 1 + (z - 1)\,e^{-\tau/T_1},
\end{aligned}
\end{equation}
i.e., populations relax exponentially toward the ground state with time constant \(T_1\) while
coherences are reduced by a factor \(e^{-\tau/(2T_1)}\).

The pure-dephasing (phase-damping) channel leaves populations unchanged and further attenuates
coherences:
\begin{equation}
\begin{aligned}
x' &= e^{-\tau/T_\varphi}\,x,\\
y' &= e^{-\tau/T_\varphi}\,y,\\
z' &= z.
\end{aligned}
\end{equation}

Combining these two effects (amplitude damping followed by pure dephasing) yields the Bloch map
in Eq.~\eqref{eq:bloch-transverse} and the relation Eq.~\eqref{eq:T2-bound} between \(T_1\),
\(T_2\), and \(T_{\varphi}\).

\section{\revised{34}{black}Nearest-neighbor double-excitation rotation implementation\endrevised{34}}\label{sec:double_imple}
\begin{equation}
    D(d) = \exp\Bigg[ i\frac{d}{8} \big( 
    \mathrm{XYXX}  + \mathrm{YXXX} + \mathrm{YYYX} +\mathrm{YYXY} 
    - \mathrm{XXYX} - \mathrm{XXXY} - \mathrm{YXYY} - \mathrm{XYYY} 
    \big) \Bigg].
\end{equation}

\begin{figure}
    \centering
    {
  \includegraphics{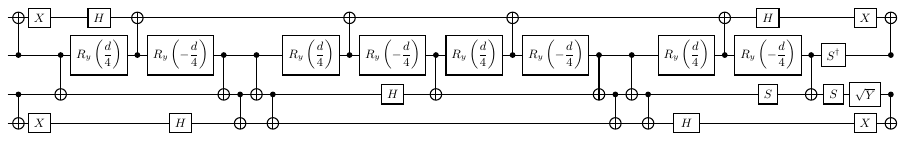} }\\
    \caption{\revised{34}{black}Nearest-neighbor, Yordanov-preserving implementation of the local double-excitation rotation used for the routed $2\times N$ FSN baseline. 
This transparent restricted-connectivity block contains 19 $\mathrm{CX}$ gates and is not claimed to be optimal among all possible $2\times2$ decompositions.\endrevised{34}}
    \label{fig:linear_yordanov}
\end{figure}

 \begingroup
\revised{35}{black}

\section{Channel-dependent error propagation in JW ladder decompositions}
\label{app:channel_dependent_noise}

The susceptibility values in Sec.~\ref{sec:numerical_study} show that the most compact circuit is not always the least
sensitive one for every Pauli noise channel. This is because the effect of a physical Pauli error
depends not only on the number of gates, but also on how the surrounding decomposition
propagates this error. Here we illustrate this point using the ladder decomposition of JW Pauli
rotations.

In the JW occupation basis, the orbital occupations are diagonal in the \(Z\) basis,
\begin{equation}
    n_j=\frac{1-Z_j}{2}.
\end{equation}
Therefore, \(Z\)-type strings preserve computational-basis occupations and do not change the
electron number of a Hartree--Fock determinant. By contrast, \(X\)- and \(Y\)-type strings can
flip occupations and may move the state away from the fixed-particle-number sector. This
distinction explains why different Pauli noise channels can affect the same circuit decomposition
differently.

Consider the Clifford part of a standard ladder decomposition. Pauli errors inserted inside the
ladder can be propagated to the input side by commuting them through the neighboring gates. We
use the identities
\begin{equation}
\begin{aligned}
    HX &= ZH,\\
    Z_1\mathrm{CX}_{12} &= \mathrm{CX}_{12}Z_1,\\
    Z_2\mathrm{CX}_{12} &= \mathrm{CX}_{12}Z_1Z_2,\\
    X_2\mathrm{CX}_{12} &= \mathrm{CX}_{12}X_2,\\
    X_1\mathrm{CX}_{12} &= \mathrm{CX}_{12}X_1X_2.
\end{aligned}
\label{eq:pauli_error_propagation_rules}
\end{equation}
The Hadamard gates exchange \(X\)- and \(Z\)-type errors, while the CX ladder spreads the
error support without changing the Pauli character except through the usual CX conjugation
rules.

\begin{figure}[bp!]
    \centering
    \includegraphics{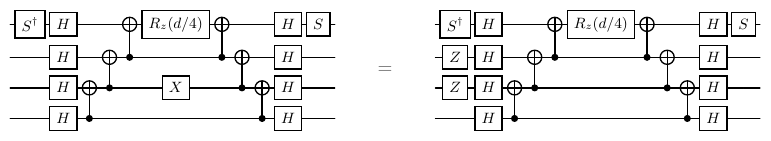}
    \caption{Propagation of an $X$-type error through a JW ladder decomposition. 
The example shows a ladder implementation of the Pauli rotation $e^{i d\,\mathrm{YXXX}/8}$. 
An $X$-type error inserted between basis-change gates can be propagated to the input side and converted into a diagonal $Z$-type string, illustrating why susceptibility can depend on the error channel and decomposition rather than only on CX count.}
    \label{fig:jw_ladder_x_error}
\end{figure}

As an example, consider the Pauli rotation
\begin{equation}
    U_{\mathrm{YXXX}}(d)=e^{i d\,\mathrm{YXXX}/8}.
\end{equation}
If an \(X\)-type error is inserted between the Hadamard layers, as in
Fig.~\ref{fig:jw_ladder_x_error}, then using
Eq.~\eqref{eq:pauli_error_propagation_rules} it can be propagated to the input side as
\begin{equation}
    E_X \circ U_{\mathrm{YXXX}}(d)
    =
    Z_1^{\alpha_1}Z_2^{\alpha_2}Z_3^{\alpha_3}Z_4^{\alpha_4}
    U_{\mathrm{YXXX}}(\pm d),
    \qquad
    \alpha_j\in\{0,1\}.
    \label{eq:x_error_to_z_string}
\end{equation}
The sign of the rotation angle depends on whether the propagated \(Z\)-string commutes or
anticommutes with the Pauli generator. The important point is that the propagated error is
diagonal in the JW occupation basis. Thus, for a state close to the Hartree--Fock determinant,
such an error mainly changes phases and does not directly change the occupation pattern.

The same \(Z\)-string can then be propagated through the remaining Pauli rotations as
\begin{equation}
    e^{i\theta P}
    Z_1^{\alpha_1}Z_2^{\alpha_2}Z_3^{\alpha_3}Z_4^{\alpha_4}
    =
    Z_1^{\alpha_1}Z_2^{\alpha_2}Z_3^{\alpha_3}Z_4^{\alpha_4}
    e^{\pm i\theta P},
    \label{eq:z_string_through_pauli_rotation}
\end{equation}
again only changing the signs of some rotation angles. This explains why JW ladder circuits can
show comparatively low susceptibility to \(X\)-type noise despite having a larger number of CX
gates.

The opposite behavior can occur for \(Z\)-type errors inserted at other positions in the same
ladder. After propagation through the basis-change gates, they may become \(X\)- or \(Y\)-type
strings, which can change occupations and therefore affect the energy more strongly. Hence the
observed noise susceptibility is decomposition- and channel-dependent, and should not be
interpreted as a function of gate count alone.

\endrevised{35}
\endgroup

\end{document}